\documentclass[]{jingdong}
\usepackage[toc,page,header]{appendix}
\PassOptionsToPackage{table}{xcolor}

\usepackage{latexsym}
\usepackage[T1]{fontenc}
\usepackage[utf8]{inputenc}
\usepackage{microtype}
\usepackage{inconsolata}
\usepackage{graphicx}
\usepackage{hyperref}       % hyperlinks
\usepackage{url}            % simple URL typesetting
\usepackage{booktabs}       % professional-quality tables
\usepackage{amsfonts}       % blackboard math symbols
\usepackage{nicefrac}       % compact symbols for 1/2, etc.
\usepackage{stackengine}
\usepackage{microtype}      % microtypography
\usepackage{colortbl}
\usepackage{xcolor}
\usepackage{amsmath}
\usepackage{amssymb}
\usepackage{amsthm}
\usepackage{mathrsfs}
\usepackage{pifont}
\usepackage{MnSymbol}
\usepackage{balance}
\usepackage{enumitem}
\usepackage{listings}
\usepackage{xcolor}
\usepackage{natbib}
\usepackage{multicol}
\AtBeginDocument{%
  \providecommand\BibTeX{{%
    \normalfont B\kern-0.5em{\scshape i\kern-0.25em b}\kern-0.8em\TeX}}}

\makeatletter
\DeclareRobustCommand\onedot{\futurelet\@let@token\@onedot}
\def\@onedot{\ifx\@let@token.\else.\null\fi}

\usepackage{setspace}
\usepackage{mathtools}

\usepackage{multirow,booktabs}
\usepackage{subcaption}

\newcommand{\owo}[1]{\textsc{OAgents}}

\definecolor{lightgreen}{RGB}{144, 238, 144} 
\definecolor{lightred}{RGB}{255, 105, 97}

\newtcolorbox{promptbox}[2][Prompt]{
colback=black!5!white,
arc=5pt, 
boxrule=0.5pt,
fonttitle=\bfseries,
title=#1, 
before upper={\small}, fontupper=\fontfamily{ptm}\selectfont,
colframe=#2, % 使用传递的参数来设定 colframe
}
\definecolor{ogreen}{RGB}{34, 139, 34}

\usepackage{algorithm}
\usepackage{algorithmicx}
\usepackage{algpseudocode}
\usepackage{wrapfig}
\usepackage{makecell}
\usepackage{bbm}

\graphicspath{{figs/}}

\usepackage{cleveref}
\theoremstyle{plain}

\theoremstyle{definition}

\theoremstyle{remark}
\usepackage{subcaption}

\usepackage[textsize=tiny]{todonotes}
\usepackage{fontawesome}

\newcommand{\ourmethod}{Echo-Infinity}
\newcommand{\memoryquery}{Memory Queries}
\newcommand{\relativerope}{Relative RoPE}

\newcommand{\sinkframe}{sink frames}
\newcommand{\localwin}{local window}

\newcommand{\fmax}{f_{\max}}

\newcommand{\Qset}{\mathcal{Q}}

\title{Echo-Infinity: Learning Evolving Memory for Real-Time Infinite Video Generation}

\newcommand{\authorlinegap}{2pt}
\newcommand{\authornl}[2][]{\addtolist[#1]{#2}{\authorlist}{\authorformat}{,\\[\authorlinegap]}}
\newcommand{\affiliationnl}[2][]{\addtolist[#1]{#2}{\affiliationlist}{\affiliationformat}{,\\[\authorlinegap]}}
\author[1]{Yuxuan Bian}
\author[2]{Zeyue Xue}
\author[3]{Songchun Zhang}
\author[4]{Shiyi Zhang}
\author[5]{Weiyang Jin}
\authornl[6]{Yaowei Li}
\author[4]{Junhao Zhuang}
\author[7]{Haoran Li}
\author[2]{Jie Huang}
\author[2]{Haoyang Huang}
\authornl[2,\dagger]{Nan Duan}
\author[1,\dagger]{Qiang Xu}

\affiliation[1]{The Chinese University of Hong Kong}
\affiliation[2]{Joy Future Academy, JD}
\affiliation[3]{The Hong Kong University of Science and Technology}
\affiliation[4]{Tsinghua University}
\affiliation[5]{The University of Hong Kong}
\affiliationnl[6]{Peking University}
\affiliation[7]{University of Science and Technology of China}

\contribution[\dagger]{Corresponding Authors}

\abstract{
We present \textbf{\textit{\ourmethod{}}}, an autoregressive~(AR) framework towards real-time infinite video generation that employs a learnable evolving memory to dynamically filter, abstract, and compress any-length history at constant cost.
Existing methods mainly curate memory with predefined KV-cache schedules, fixed-ratio heuristic compression, or inference-time RoPE adaptation.
These designs inevitably lose historical information and amplify compounding errors due to their limited cache window and ignorance of autoregressive generation noise.
Inspired by human memory consolidation, \textit{\ourmethod{}} replaces handcrafted memory curation with learnable \textbf{\textit{Memory Queries}}, which are updated by attention and a gating mechanism when past frames are evicted from the \localwin{}.
The queries are optimized end-to-end with the video diffusion transformers~(DiTs), forming an evolving memory that supports arbitrary compression ratios with constant computation independent of video length.
They also act as a generalizable generation prior, improving quality even when only the optimized initial state is used.
We further introduce \textbf{\textit{Unified Relative RoPE} Recipe}, which anchors the sink frames to start from id $0$ and lets the newest frame id grow at most to the DiTs' pretrained maximum temporal RoPE id $\fmax$ throughout training and inference, freeing the model from the finite RoPE constraint and closing the train-test RoPE extrapolation gap.
In long and short video generation, \textit{\ourmethod{}} achieves state-of-the-art performance, and, to our knowledge, demonstrates promising \textbf{$\textbf{24-hour}$ ($\textbf{>1.3 M}$ frames) real-time rollouts} for the first time, suggesting a practical path toward infinite video generation.
}

\date{\today}
\checkdata[Project Page]{\url{https://echo-team-joy-future-academy-jd.github.io/Echo-Infinity/}}
\checkdata[Code]{\url{https://github.com/Echo-Team-Joy-Future-Academy-JD/Echo-Infinity}}

\begin{document}
\maketitle

% \vspace{-3pt}
\section{Introduction}
\label{sec:intro}
% \vspace{-3pt}

Modern video diffusion transformers~(DiTs) have advanced high-quality video generation~\citep{wan2025wan,ltx2}.
Autoregressive~(AR) DiTs further enable real-time streaming~\citep{selfforcing,longlive}, but long horizons expose two bottlenecks: prohibitive memory caused by unbounded history key-value~(KV) caches and temporal rotary positional embedding~(RoPE) indices exceeding the training range, causing degradation and overflow.

\begin{figure}[t]
  \centering
  % \vspace{-26pt}
  \includegraphics[width=\linewidth]{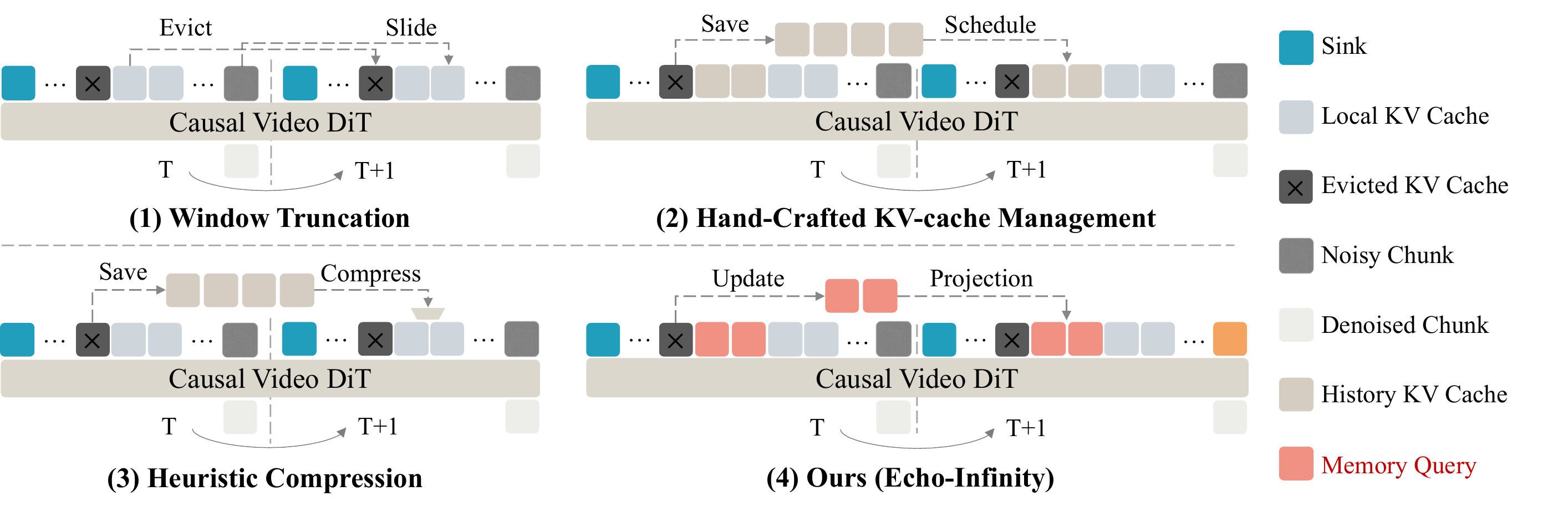}
  % \vspace{-10pt}
  \caption{\textbf{Memory Mechanisms in Long Video Generation.} Different from handcrafted curation, \ourmethod{} proposes end-to-end trainable memory queries as the evolving long-term memory.}
  \label{fig:related-taxonomy}
  % \vspace{-10pt}
\end{figure}

To address the unbounded KV-cache challenge, most existing work adopts passive, predefined strategies, which fall into three families (see Fig.~\ref{fig:related-taxonomy}).
\textbf{(1)~Window Truncation:} methods~\citep{longlive,rollingforcing} keep only a bounded local window with a few \sinkframe{}, controlling memory cost but discarding distant history.
\textbf{(2)~Hand-Crafted KV-cache Management:} methods~\citep{memflow,contextforcing,packforcing,groundedforcing} retain selected evicted KVs by offline rules, but remain tied to fixed cache budgets and ignore the compounding error within history.
\textbf{(3)~Heuristic Compression:} methods~\citep{memorizeandgenerate,lovic,pretrainingframepreservation} compress history into compact representations, yet often rely on separate objectives or predefined compression ratios/schedules rather than an end-to-end learned fixed-capacity memory.

\begin{figure}[t]
  \centering
  % 
  % \vspace{-16pt}
  % 
  \includegraphics[width=\linewidth]{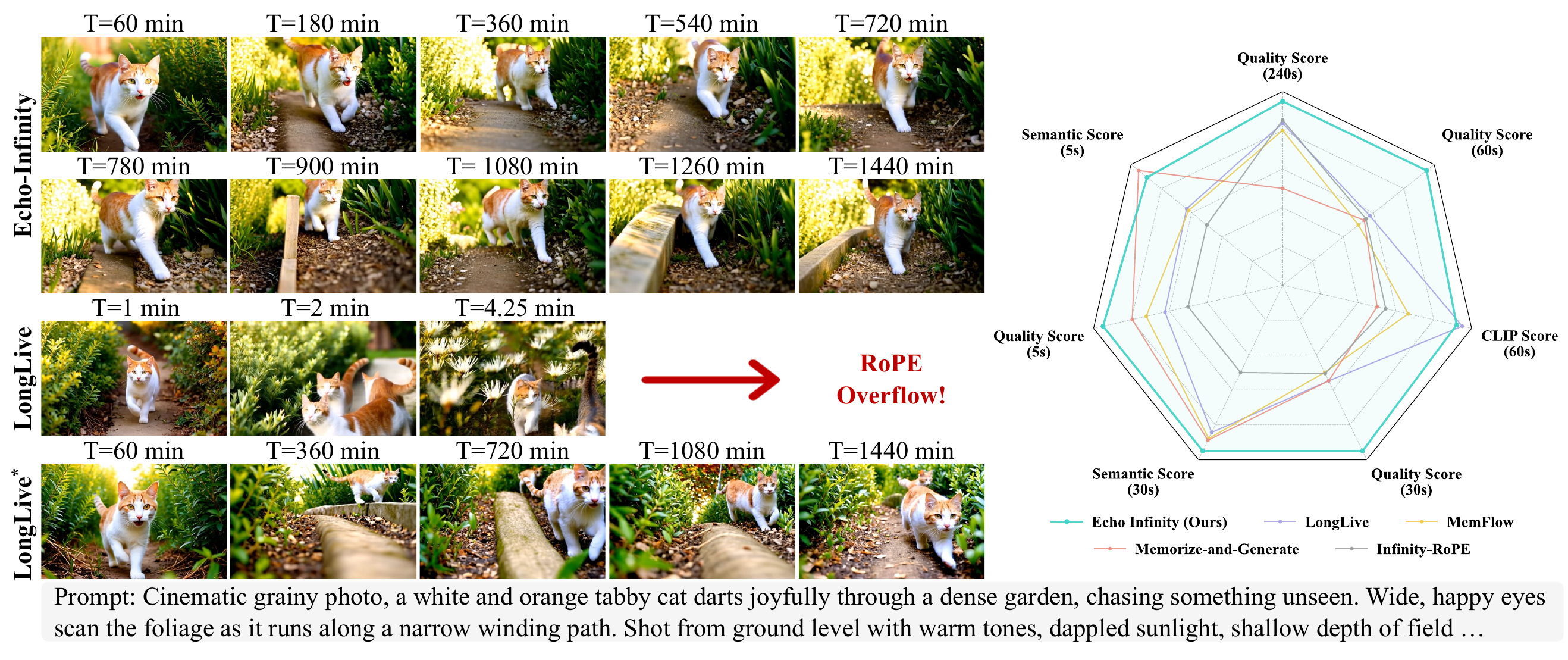}
  % 
  % \vspace{-10pt}
  % 
  \caption{
    \textbf{\ourmethod{} at a glance.}
    Left: \ourmethod{} can generate extremely long videos in real-time over \textbf{$\textbf{24}$ hours ($>\textbf{1.3 M}$ frames)},
    while LongLive, constrained by the absolute RoPE, degrades dramatically for $4.25$ minutes and then overflows. 
    Even equipped with the relative RoPE in inference ({LongLive}$^{*}$), its consistency still declines due to unsolved train-test RoPE extrapolation and limited long-term memory.
    \textbf{Right}: \ourmethod{} achieves overall state-of-the-art performance across various video generation benchmarks.
    }
    % \vspace{-10pt}
  \label{fig:teaser}
\end{figure}

Human memory suggests a different design.
Cognitive neuroscience~\citep{gabrieli1998cognitive} describes memory as hierarchical: \textbf{Fresh perceptions first enter a working buffer, then are selectively filtered, abstracted, and compressed into a compact long-term store.}
These observations suggest that long-horizon memory depends less on storing every past token and more on a compact, evolving state.
We therefore make memory end-to-end learnable for long video generation by introducing learnable \textbf{\memoryquery{}}, jointly optimized with a pretrained DiT, as a long-term memory state.
Whenever frames are evicted from the local window, these queries are refreshed by a cross-attention update followed by a sigmoid-gated residual, enabling~~\textbf{filtering}, \textbf{abstraction}, and~\textbf{compression} akin to human memory consolidation, and provide the~\textbf{``history echo''}~(see Fig. \ref{fig:related-taxonomy}).

Beyond memory, the risks of temporal RoPE extrapolation and overflow still exist.
Concurrent methods~\citep{infinityrope,memrope} mitigate overflow only at inference time by rotating previous caches backward once the newest id reaches the pretrained maximum $\fmax$.
Because their models are still trained with absolute RoPE, the train-test mismatch remains.
We introduce a \textbf{Unified \relativerope{} Recipe} that applies the same bounded relative-RoPE schedule during both training and inference, keeping all active temporal ids within the pretrained range.

We present \ourmethod{}, a real-time autoregressive video generation framework toward infinite-horizon generation with an end-to-end optimized evolving memory state at constant cost~(Fig.~\ref{fig:teaser}).
At its core are the \textbf{\textit{Memory Queries}}, a set of trainable tokens that preserves generation-relevant information as a compact evolving state alongside local KV caches and sink frames.
When frames are evicted from the local window, the queries are refreshed by an \emph{attention update} over evicted KV caches to extract relevant information, followed by a \emph{gated residual} that controls memory overwriting, enabling filtering, abstraction, and compression of any-length history at constant cost.
To avoid temporal RoPE overflow and train-test mismatch, we employ \textbf{Unified \relativerope{} Recipe}: it anchors the \sinkframe{} at id $0$, lets the newest frame grow up to the pretrained maximum $\fmax$, and rotates older frames backward once $\fmax$ is reached, keeping all temporal RoPE ids within the pretrained range during both training and inference.%

We evaluate \ourmethod{} on long~(30 s / 60 s / 240 s) and short~(5 s) video benchmarks and obtain overall state-of-the-art performance~(see \S\ref{sec:exp-long}, \S\ref{sec:exp-interactive}, and \S\ref{sec:exp-short}).
In short-video generation, the optimized memory queries can also serve as a generalizable video generation prior even when memory updates are manually disabled.
Furthermore, \ourmethod{} demonstrates promising \textbf{day-scale real-time generation results over $\textbf{24}$ hours and $\textbf{>1.3M}$ frames at $\textbf{18.5}$ FPS} on a single NVIDIA H100, with only $10.6\%$ throughput overhead over a memory-free baseline.
To summarize, our contributions are:
\begin{itemize}[
% leftmargin=*,
% labelindent=0pt,
% itemsep=0.15em,
% topsep=0.25em,
% parsep=0pt,
% partopsep=0pt
]

\item We propose \textbf{\ourmethod{}}, an autoregressive framework towards real-time long-horizon video generation, replacing handcrafted memory curation with end-to-end trainable \textbf{\memoryquery{}} that filter, abstract, and compress arbitrary-length history at constant cost.

\item We employ \textbf{Unified \relativerope{} Recipe}, which keeps every active temporal RoPE id within the trained range throughout training and inference, avoiding RoPE train-test extrapolation or overflow.
\item We achieve generalizable and state-of-the-art performance on long, short, and interactive video generation. We further provide the first demonstration of promising results over $\textbf{24-hour}$ \textbf{real-time video generation} \textbf{($\textbf{>1.3M}$ frames)}, paving the way toward infinite video generation.
\end{itemize}

% \vspace{-5pt}
\section{Related Work}
\label{sec:related}

% \vspace{-5pt}
\subsection{Long Video Generation}
% \vspace{-5pt}

Modern video diffusion models~\citep{cogvideox,hunyuan-video,wan2025wan,ltxvideo,moviegen,skyreelsv2} mostly use bidirectional DiTs, whose quadratic attention cost and bidirectional generation make them unsuitable for streaming long videos.
Recent autoregressive methods~\citep{nova,pyramidflow,magi1}, together with Diffusion Forcing~\citep{diffusion-forcing}, DMD distillation~\citep{improved-dmd,dmd}, and Self-Forcing~\citep{causvid,selfforcing}, push video generation toward minute-scale and real-time streaming~\citep{longlive,rollingforcing,selfforcingpp,framepack}.
\ourmethod{} targets the remaining memory and positional bottlenecks for longer-horizon AR generation.

% \vspace{-5pt}
\subsection{Memory Mechanisms in Long Video Generation}
% \vspace{-5pt}

Current methods mainly address the unbounded KV cache based on specially designed memory mechanisms:
\noindent\textbf{(a) Window Truncation.}\quad
These methods retain a bounded local window plus a few \sinkframe{} and discard others, leading to inevitable history loss~\citep{longlive,rollingforcing}.
\noindent\textbf{(b) Hand-Crafted KV-cache Management.}\quad
A second family augments the local window with rule-based scheduling that decides which evicted KVs to keep~\citep{memflow,contextforcing,relaxforcing,packforcing,anchorforcing,groundedforcing}.
For example, MemFlow~\citep{memflow} curates a memory bank by textual retrieval.
However, these rules are tuned offline and remain bound by the given window length.
\noindent\textbf{(c) Heuristic Compression.}\quad
Rather than selecting evicted KVs, a third family replaces the evicted history with compressed representations~\citep{memorizeandgenerate,lovic,framepack,pretrainingframepreservation}.
Memorize-and-Generate~\citep{memorizeandgenerate} decouples memory compression and frame generation by compressing historical information into compact KVs via reconstruction, which are consumed by a separate generator.
%
% LoViC~\citep{lovic} compresses long video-text context into latent context tokens with a separately trained autoencoder and feeds them to a DiT decoder.
%
These methods improve the temporal context, but their compressed states are still tied to predefined compression ratios, compression schedules, or separate reconstruction/compression stages.
VideoSSM~\citep{videossm} further introduces an SSM-based evolving global memory, but requires architecture-level state-space modules.
In contrast, \ourmethod{} recurrently consolidates evicted causal KVs into a persistent memory state used directly as a plug-in KV source, learning what history to preserve end-to-end under the long-video generation objective while keeping the active memory footprint independent of sequence length.

% \vspace{-6pt}
\subsection{Rotary Positional Embedding for Long Video Generation}
\label{sec:related-rope}
% \vspace{-6pt}

Modern video DiTs apply 3D rotary positional embeddings~(RoPE) independently along temporal, height, and width axes~\citep{rope}.
During autoregressive rollouts, the temporal index quickly exceeds the pretraining range~(e.g., maximum $20$ for Wan-2.1~\citep{wan2025wan}), causing quality collapse and overflow.
Concurrent training-free methods~\citep{infinityrope,deepforcing,memrope} mitigate overflow at inference time with relative RoPE, but leave train-test inconsistency unresolved.
\ourmethod{} eliminates this mismatch with a unified relative RoPE schedule that keeps every active temporal id within the same range in both training and inference.

\begin{figure}[t]
  \centering
  % \vspace{-25pt}
  \includegraphics[width=\linewidth]{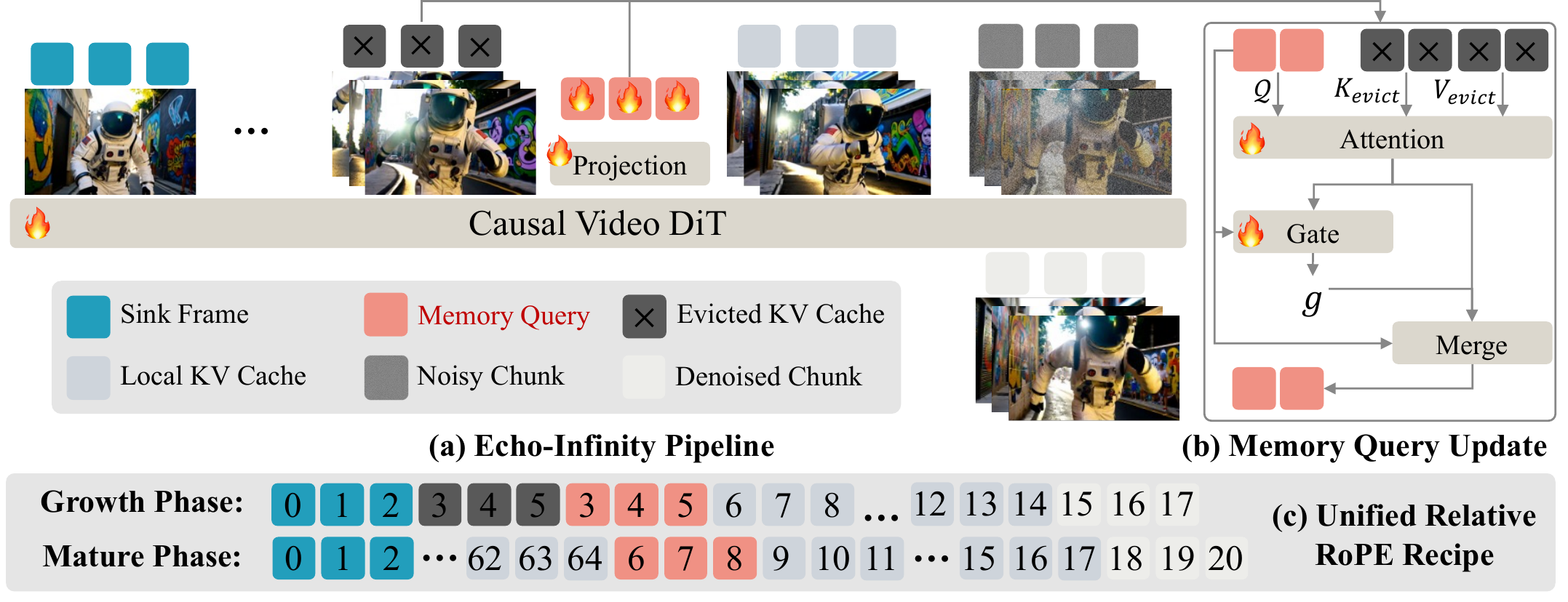}
  % \vspace{-10pt}
  % 
  \caption{
  \textbf{Overview.} 
  \textbf{\ourmethod{}} introduces end-to-end trainable \textbf{Memory Queries} that filter, abstract, and compress evicted history KV caches through attention and gating, enabling evolving compression of arbitrarily long histories.
  To avoid temporal RoPE extrapolation during inference or even overflow, \ourmethod{} uses \textbf{Relative RoPE} throughout training and inference, which anchors the \sinkframe{} to start from id $0$ and caps the newest id by the backbone's pretrained maximum id $\fmax$~(e.g., $\fmax=20$ for Wan2.1-1.3B~\citep{wan2025wan}), closing the RoPE extrapolation gap.
  }
  % \vspace{-10pt}
  % 
  \label{fig:method-overview}
\end{figure}

% \vspace{-5pt}
\section{Method}
\label{sec:method}
% \vspace{-5pt}

\subsection{Preliminaries}
\label{sec:method-prelim}

% \vspace{-5pt}
\paragraph{Distribution Matching Distillation~(DMD).}
DMD~\citep{improved-dmd,dmd} distills a multi-step diffusion model into a few-step generator $G_\theta$ by matching the generated distribution $p_{\mathrm{fake}}(\mathbf{x})$ to the real distribution $p_{\mathrm{real}}(\mathbf{x})$ through reverse KL minimization over diffusion timesteps:
\begingroup
\setlength{\abovedisplayskip}{4pt}
\setlength{\belowdisplayskip}{4pt}
\setlength{\abovedisplayshortskip}{4pt}
\setlength{\belowdisplayshortskip}{4pt}
\setlength{\jot}{4pt}
\begin{equation}
\begin{split}
\nabla_\theta \mathcal{L}_{\mathrm{DMD}}
&\triangleq \mathbb{E}_{t}\!\left(\nabla_\theta\,\mathbb{D}_{\mathrm{KL}}\!\left(p_{\mathrm{fake},t}(\mathbf{x}_{t})\,\big\|\,p_{\mathrm{real},t}(\mathbf{x}_{t})\right)\right)\\
&\approx -\mathbb{E}_{t}\!\left(\int \!\big(s_{\mathrm{real}}(\Psi(G_\theta(\boldsymbol{\epsilon}),t),t) - s_{\mathrm{fake}}(\Psi(G_\theta(\boldsymbol{\epsilon}),t),t)\big)\,\frac{\mathrm{d}G_\theta(\boldsymbol{\epsilon})}{\mathrm{d}\theta}\,\mathrm{d}\boldsymbol{\epsilon}\right),
\end{split}
\label{eq:dmd}
\end{equation}
\endgroup
where $\boldsymbol{\epsilon} \sim \mathcal{N}(\mathbf{0}, \mathbf{I})$ and $\Psi$ denotes forward diffusion at timestep $t$.
The score function is as follows:
%
% \vspace{-5pt}
% 
\begingroup
\setlength{\abovedisplayskip}{4pt}
\setlength{\belowdisplayskip}{4pt}
\setlength{\abovedisplayshortskip}{4pt}
\setlength{\belowdisplayshortskip}{4pt}
\setlength{\jot}{4pt}
\begin{equation}
s_{\mathrm{real}}(\mathbf{x}_{t}, t) = \nabla_{\mathbf{x}_{t}}\log p_{\mathrm{real},t}(\mathbf{x}_{t}) = -\frac{\mathbf{x}_{t} - \alpha_{t}\,\mu_{\mathrm{real}}(\mathbf{x}_{t}, t)}{\sigma_{t}^{2}},
\label{eq:score}
\end{equation}
\endgroup
%
% \vspace{-5pt}
% 
where $\mu_{\mathrm{real}}$ is the denoised estimate and $\alpha_{t}, \sigma_{t}$ are noise-schedule parameters~\citep{ho2020denoising,karras2022elucidating}.
In DMD, the pre-trained $\mu_{\mathrm{real}}$ is frozen, while $\mu_{\mathrm{fake}}$ is learned from generator samples.

% \vspace{-10pt}
\paragraph{KV Cache and RoPE.}
In causal Video DiTs, at layer $l$, stored keys and values ${K^l_{f'}, V^l_{f'}}$ for $f' \le f$ form a KV cache that grows linearly with $f$ and dominates memory in long inference.
We split the cache into three tiers: a \sinkframe{} block of size $N_S$ as a persistent global anchor~\citep{streamingllm,longlive}, a \localwin{} of size $N_W$ that keeps recent KV cache, and trainable and evolving memory queries that summarize evicted history.
DiTs also employ 3-D RoPE~\citep{dit,rope}.
Each frame has a temporal id $f \in \{0,\dots,\fmax,\fmax+1,\dots\}$, which rotates queries and keys with phase $\Theta(f,h,w)$, where $\fmax$ is the maximum index during training.
% \vspace{-5pt}
\subsection{Overall Framework}
\label{sec:method-overview}
% \vspace{-5pt}

Following previous works~\citep{selfforcing,longlive}, \ourmethod{} builds on a two-stage DMD training strategy: standard-tuning~(5s standard video generation without prompt switching), and the streaming long-tuning~(60s long video generation with one prompt switching).
We further introduce end-to-end trainable memory queries that support arbitrary compression ratios for the increasing KV cache, mimicking the filtering, abstraction, and compression of human long-term memory consolidation throughout the two stages.

% \vspace{-10pt}
\paragraph{Three-tier KV organization mirrors human memory.}
Following the cognitive-science motivation of \S\ref{sec:intro}, \ourmethod{} structures every layer's KV cache into three compartments that mirror the human memory hierarchy.
The \sinkframe{} block of size $N_S$  acts as a persistent global anchor, analogous to basic defining memories.
The \localwin{} of size $N_W$ is a short-term working buffer for the most recent frames.
A learnable and evolving memory query set $\Qset = \{q_1, \dots, q_{N_Q}\}$ stores the filtered, abstracted, and compressed long-term memories of all evicted history.
As shown in Fig.~\ref{fig:method-overview}, at every chunk-generation step (chunk size $B$, current frame index $f_{*}$), the \sinkframe{} / memory queries / \localwin{} and the current chunk tile the active RoPE temporal ids over $[0, \fmax]$:
\begingroup
\setlength{\abovedisplayskip}{6pt}
\setlength{\belowdisplayskip}{6pt}
\setlength{\abovedisplayshortskip}{6pt}
\setlength{\belowdisplayshortskip}{6pt}
\setlength{\jot}{6pt}
\begin{equation}
\resizebox{0.948\linewidth}{!}{$\displaystyle
\underbrace{[\,0,\, \dots,\, N_S{-}1\,]}_{\mathrm{Sink} \;(N_S\,\mathrm{frames})}
\;\Vert\;
\underbrace{[\,r^{\mathrm{cur}}_{\mathrm{start}} - N_W -N_Q,\, \dots,\, r^{\mathrm{cur}}_{\mathrm{start}} - N_W-1\,]}_{\mathrm{Memory~Queries}~\Qset \;(N_Q\,\mathrm{frames})}
\;\Vert\;
\underbrace{[\,\, r^{\mathrm{cur}}_{\mathrm{start}} - N_W,\, \dots,\, r^{\mathrm{cur}}_{\mathrm{start}} - 1\,]}_{\mathrm{Local\,Window} \;(N_W\,\mathrm{frames})}
\;\Vert\;
\underbrace{[\,r^{\mathrm{cur}}_{\mathrm{start}},\, \dots,\, r^{\mathrm{cur}}_{\mathrm{end}}\,]}_{\mathrm{Current\;Chunk} \;(B\,\mathrm{frames})},
$}
\label{eq:rrope-layout}
\end{equation}
\endgroup
where $r^{\mathrm{cur}}{\mathrm{end}} = \min(f{*} + B - 1, \fmax)$, with $\fmax$ the maximum pretraining id. The remaining intervals are derived backward from $r^{\mathrm{cur}}_{\mathrm{end}}$, and $\Qset$ appears only after KV eviction. 

% \vspace{-10pt}
\paragraph{Generation Process.}
At iteration $m$ for chunk $\mathbf{X}_m=\{{\mathbf{x}_{f_{*}}, \dots, \mathbf{x}_{f_{*}+B-1}}\}$, each transformer layer $l$ has three steps.
(i) \textbf{Three-tier Attention}: $\mathbf{X}_m$ attends to $K^l_{\mathrm{sink}} \cup K_{\Qset} \cup K^l_{\mathrm{local}}$ and the corresponding values, with memory queries $\Qset$ shared across layers~(detailed in Fig. \ref{fig:method-overview} (a)).
(ii) \textbf{Memory Update}: the new chunk's KVs enter the local window, while evicted KVs are routed to $\Qset$ for filtering, abstraction, and compression~(detailed in \S\ref{sec:method-memquery}, Fig. \ref{fig:method-overview} (b)).
(iii) \textbf{RoPE Rescheduling}: all caches are reassigned ids under the relative RoPE schedule (detailed in \S\ref{sec:method-relrope}, Fig. \ref{fig:method-overview} (c)), keeping every active id within $[0, \fmax]$.

% \vspace{-6pt}
\subsection{Memory Queries}
\label{sec:method-memquery}
% \vspace{-6pt}

% \paragraph{Learnable memory queries.}
The core of \ourmethod{} is a compact set of learnable memory queries $\Qset \in \mathbb{R}^{1 \times N_Q \cdot S \times d}$ ($S$ is the latent token number of a video frame), whose representation is optimized end-to-end during the DMD-based autoregressive video generation training~\citep{causalforcing,longlive}.

% \vspace{-10pt}
\paragraph{Update on eviction.}
As in Fig. \ref{fig:method-overview} (b), when the local window slides forward and a set of frames is evicted from the local KV cache window, we feed the corresponding KV cache $\{K_{\mathrm{evict}} \in \mathbb{R}^{1 \times B \cdot S \times d}, V_{\mathrm{evict}} \in \mathbb{R}^{1 \times B \cdot S \times d}\}$ of last layer, whose representations are closest to the pixel output and thus suitable for layer-shared memory queries, into an encoder $\mathrm{Enc}$ with $L_{\mathrm{enc}}$ layers of cross-attention to refresh $\Qset$, followed by a sigmoid-gated residual that updates the state:
\begingroup
\setlength{\abovedisplayskip}{6pt}
\setlength{\belowdisplayskip}{6pt}
\setlength{\abovedisplayshortskip}{6pt}
\setlength{\belowdisplayshortskip}{6pt}
\setlength{\jot}{6pt}
\begin{equation}
\resizebox{0.94\linewidth}{!}{$\displaystyle
\widetilde{\Qset} \;=\; \mathrm{Enc}\!\left(\Qset;\;K_{\mathrm{evict}};\;V_{\mathrm{evict}}\right),\qquad
\mathbf{g} \;=\; \sigma([\Qset;\,\widetilde{\Qset}]\,W_{\mathrm{gate}}),\qquad
\Qset \;\leftarrow\; \mathbf{g}\odot\Qset \;+\; (1-\mathbf{g})\odot\widetilde{\Qset},
$}
\label{eq:memq-update}
\end{equation}
\endgroup
where $[\cdot]$ denotes concatenation and $W_{\mathrm{gate}} \in \mathbb{R}^{2d \times d}$.
End-to-end training guides the cross-attention and gate to select, abstract, and compress the most useful information from the evicted KV cache into the current $\Qset$, akin to the consolidation mechanisms of long-term human memory~\citep{gabrieli1998cognitive}.

% \vspace{-5pt}
\paragraph{Injection and end-to-end optimization.}
At each layer $l$, $\Qset$ is projected into $K_{\Qset}/V_{\Qset}$ by two linear maps $W_k^{\Qset} \in \mathbb{R}^{d \times d}, W_v^{\Qset} \in \mathbb{R}^{d \times d}$ (shared across layers), and concatenated with the sink and local KVs before causal attention, serving as a plug-in key/value source without changing the backbone structure.
In stage 1 training~(5 s), memory parameters $\{\Qset, \mathrm{Enc}, W_{\mathrm{gate}}, W_k^{\mathcal{Q}}, W_v^{\mathcal{Q}}\}$ are jointly optimized with $G_\theta$ under Eq.~\ref{eq:dmd}, with gradients propagated through all memory updates to the shared $\Qset$.
For computational efficiency, during stage-2 long tuning, we detach the $\Qset$ state and cached sink/local KVs only across 5-s sub-clip boundaries, while continuing to optimize trainable model and memory parameters within each sub-clip.
We find that stage-1 training distills video priors into $\Qset$: even when Eq.~\ref{eq:memq-update} is disabled in short-video generation, the optimized $\Qset$ still benefits generation~(\S\ref{sec:exp-short}).
Notably, since $\Qset$~($N_Q$) and the evicted KV cache size~($B$) are fixed, the extra memory-query cost remains constant throughout generation and does not grow with video length.

% \vspace{-6pt}
\subsection{Unified Relative RoPE Recipe}
\label{sec:method-relrope}
% \vspace{-6pt}

\begin{wrapfigure}{r}{0.48\textwidth}
  \vspace{-40pt}
  \centering
  \scalebox{0.75}{%
  \begin{minipage}[t]{1.33\linewidth}
    \begin{algorithm}[H]
      \caption{\ourmethod{}: single chunk step.}
      \small
      \begin{algorithmic}[1]
        \Require Per-layer cache $\mathcal{C}{=}(K_{\mathrm{sink}}, V_{\mathrm{sink}}, K_{\mathrm{local}}, V_{\mathrm{local}}, K_{\mathrm{cur}}, V_{\mathrm{cur}})$
        \Require Memory $\Qset$ with flag $\mathrm{has\_history}$
        \Require Frame index $f_{*}$, chunk size $B$, window $N_W$, prompt $P$
        \Ensure Chunk $\mathbf{x}_{[f_{*}:f_{*}+B-1]}$; updated $\mathcal{C}, \Qset, \mathrm{has\_history}$
                \State $\mathbf{r} \gets \textsc{CompRoPE}(f_{*}, B, |\mathrm{sink}|, |\Qset|, |\mathrm{local}|, \fmax)$ \Comment{Eq.~\ref{eq:rrope-layout}}
        \Statex \hspace{\algorithmicindent}\textbf{where } $\mathbf{r}=(r^{\mathrm{sink}}, r^{\Qset}, r^{\mathrm{local}}, r^{\mathrm{cur}})$
        \If{$\mathrm{has\_history}$}
          \State $K_{\mathcal{Q}} \gets \textsc{RoPE}(W_k^{\mathcal{Q}}\Qset, r^{\mathcal{Q}})$, $V_{\mathcal{Q}} \gets W_v^{\mathcal{Q}}\Qset$
        \Else
          \State $K_{\mathcal{Q}}, V_{\mathcal{Q}} \gets \emptyset$
        \EndIf
        \For{$l = 1, \dots, L$}
          \State Apply $r^{\mathrm{sink}}, r^{\mathrm{local}}, r^{\mathrm{cur}}$ to $K^l_{\mathrm{sink}}, K^l_{\mathrm{local}}, K^l_{\mathrm{cur}}, Q^l_{\mathrm{cur}}$
          
          \State $K^l_{\mathrm{all}} \gets [K^l_{\mathrm{sink}}; K_{\mathcal{Q}}; K^l_{\mathrm{local}}; K^l_{\mathrm{cur}}]$, 
          \State $V_{\mathrm{all}} \gets [V^l_{\mathrm{sink}}; V_{\mathcal{Q}}; V^l_{\mathrm{local}}; V^l_{\mathrm{cur}}]$
          \State $\mathbf{x}^{(l)} \gets \textsc{CausalAttn}(Q_{cur}^{l}, K^l_{\mathrm{all}}, V^l_{\mathrm{all}})$
        \EndFor
        \State Sample $\mathbf{x}_{[f_{*}:f_{*}+B-1]}$ from $\mathbf{x}^{(L)}$ conditioned on $P$
        \State Append new $K, V$ to $K_{\mathrm{local}}, V_{\mathrm{local}}$; evict if $|\mathrm{local}| > N_W$
        \If{any frame evicted}
          \State $\Qset \gets \textsc{Update}(\Qset;  K_{\mathrm{evict}}, V_{\mathrm{evict}})$ \Comment{Eqs.~\ref{eq:memq-update}}
          \State $\mathrm{has\_history} \gets \text{True}$
        \EndIf
        
        \State \Return $\mathbf{x}, \mathcal{C}, \Qset, \mathrm{has\_history}$
      \end{algorithmic}
      \label{alg:echoforcing}
    \end{algorithm}
  \end{minipage}%
  }
  \vspace{-2em}
  \end{wrapfigure}

\paragraph{Train-Test Mismatch.}
As DiT's temporal RoPE id $r$ is trained only on $\{0, 1, \dots, \fmax\}$, naively generating beyond $\fmax$ introduces unseen ids outside the trained region, leading to rapid degradation and even overflow.
Our unified relative RoPE recipe tackles these by combining Eq.~\ref{eq:rrope-layout} with the below schedule throughout training and inference.

% \vspace{-8pt}
\paragraph{\relativerope{} Schedule.}
As seen in Fig.~\ref{fig:method-overview} (c), the schedule has two phases.
(1) In the \emph{growth phase}, $r^{\mathrm{cur}}_{\mathrm{end}}$ increases from $|\mathrm{sink}|$ to $\fmax$ as chunks are generated.
(2) In the \emph{mature phase}, $r^{\mathrm{cur}}_{\mathrm{end}}$ stays at $\fmax$.
For each new chunk, all non-sink ids rotate backward from $r^{\mathrm{cur}}_{\mathrm{end}}$, equivalently shifting older frames forward by one unit, while the sink frames remain to start from $0$.
Thus, all temporal ids stay within $[0, \fmax]$ without overflow risks and train-test mismatch.

Algorithm~\ref{alg:echoforcing} summarizes one chunk-generation step of \ourmethod{}, combining the three-tier KV cache in \S\ref{sec:method-overview}, the memory-query update in \S\ref{sec:method-memquery}, and the \relativerope{} schedule in \S\ref{sec:method-relrope}.

\begin{table*}[t]
    \centering
    \caption{\textbf{Single-Prompt 30s / 240s Long Video Evaluation} on VBench-Long~\citep{huang2025vbench++} / MovieGen~\citep{moviegen}. \ourmethod{} outperforms baselines in generation scores and user preference, benefiting from end-to-end learned memory queries and the unified relative RoPE recipe.}
    \label{tab:long-merged}
    \small
    \setlength{\tabcolsep}{2pt}
    \resizebox{\linewidth}{!}{%
    \begin{tabular}{l c c ccc cc}
    \toprule
    \multirow{2}{*}{\textbf{Model}} &
    \multirow{2}{*}{\textbf{\#Params}} &
    \multirow{2}{*}{\makecell{\textbf{Throughput} \\ \textbf{(FPS)} $\uparrow$}} &
    \multicolumn{3}{c}{\textbf{Evaluation scores on $30\,$s} $\uparrow$} &
    \multicolumn{2}{c}{\textbf{Evaluation scores on $240\,$s} $\uparrow$} \\
    \cmidrule(lr){4-6} \cmidrule(lr){7-8}
    & & & Quality & Semantic & User Preference~($\%$) & Quality & User Preference~($\%$) \\
    \midrule
    LongLive~\citep{longlive}                         & 1.3B & \underline{20.7} & 83.59 & 80.28 & 10.47 & 79.79 & 6.13 \\
    MemFlow~\citep{memflow}                           & 1.3B & 18.7 & 83.35 & 80.85 & 10.13 & 79.31 & 5.93 \\
    Memorize-and-Generate~\citep{memorizeandgenerate} & 1.3B & \textbf{21.7} & \underline{83.69} & \underline{81.01} & \underline{14.73} & 75.49 & 2.13 \\
    $\infty$-RoPE~\citep{infinityrope}                & 1.3B & 17.0 & 83.38 & 74.67 & 5.13 & \underline{79.99} & \underline{14.13} \\
    \midrule
    % \rowcolor{yellow!20}
    \textbf{\ourmethod{} (Ours)}                      & 1.3B & 18.5 & \textbf{85.61} & \textbf{82.01} & \textbf{59.53} & \textbf{81.23} & \textbf{71.67} \\
    \bottomrule
    \end{tabular}}
    % \vspace{-5pt}
\end{table*}

% \vspace{-3pt}
\section{Experiments}
\label{sec:exp}

\begin{figure}[t]
  \centering
  % \vspace{-15pt}
  \includegraphics[width=\linewidth]{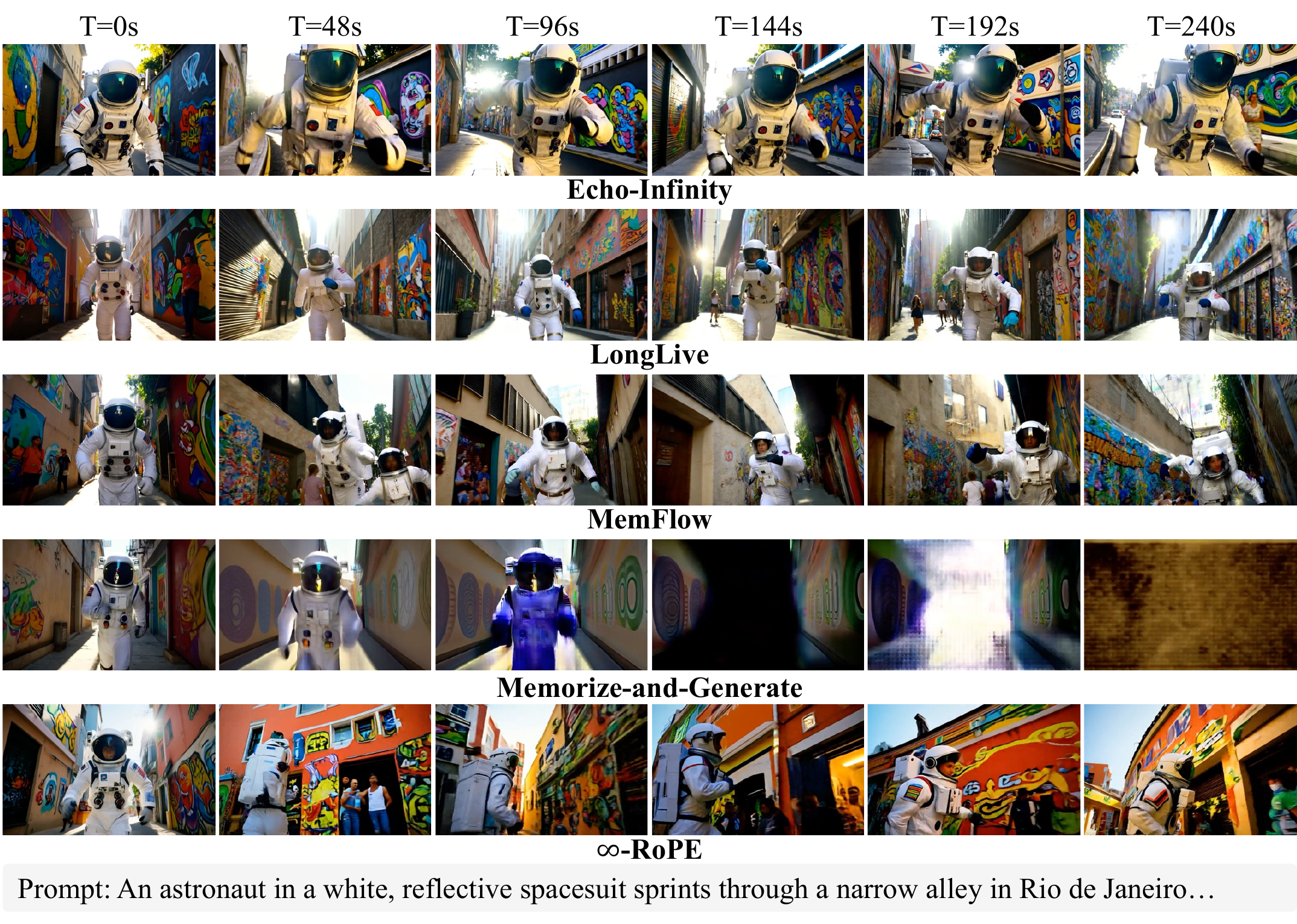}
  % \vspace{-20pt}
  \caption{\textbf{$\textbf{240}\,$s Long Video Generation Result Comparison.} More visualizations are in the supplementary.}
  % \vspace{-5pt}
  \label{fig:exp-vis-240s}
\end{figure}

\begin{table}[!t]
    \centering
    \caption{\textbf{Multi-Prompt $\textbf{60}\,$s Interactive Evaluation} on MemFlow~\citep{memflow}'s interactive video generation benchmark.}
    % \vspace{-5pt}
    \label{tab:interactive-60s}
    \small
    \renewcommand{\arraystretch}{1.08}
    \resizebox{\linewidth}{!}{%
    \begin{tabular}{@{}lccccccc@{}}
    \toprule
    \multirow{2}{*}{\textbf{Method}} &
    \multirow{2}{*}{\shortstack{\textbf{Quality}\\\textbf{Score} $\uparrow$}} &
    \multicolumn{6}{c}{\textbf{CLIP Score} $\uparrow$} \\
    \cmidrule(lr){3-8}
    & & 0--10\,s & 10--20\,s & 20--30\,s & 30--40\,s & 40--50\,s & 50--60\,s \\
    \midrule
    LongLive~\citep{longlive}                         & 79.38 & \underline{34.08} & 32.09 & \textbf{32.03} & \textbf{31.55} & \textbf{30.88} & \underline{30.49} \\
    MemFlow~\citep{memflow}                           & \underline{79.91} & 33.48 & 31.94 & 31.95 & 30.87 & 30.53 & 30.23 \\
    Memorize-and-Generate~\citep{memorizeandgenerate} & 79.15 & 33.58 & 31.43 & 31.14 & 30.65 & 30.48 & 30.27 \\
    $\infty$-RoPE~\citep{infinityrope}                & 79.22 & 33.15 & \textbf{32.47} & 31.41 & 30.46 & 30.29 & 30.17 \\
    \midrule
    % \rowcolor{yellow!20}
    \textbf{\ourmethod{} (Ours)}                      & \textbf{81.71} & \textbf{34.10} & \underline{32.42} & \underline{31.99} & \underline{31.18} & \underline{30.83} & \textbf{30.74} \\
    \bottomrule
    \end{tabular}}
    % \vspace{-18pt}
\end{table}

% \vspace{-3pt}
\subsection{Implementation}
\label{sec:exp-impl}
% \vspace{-3pt}

We implement \ourmethod{} on Wan2.1-T2V-1.3B~\citep{wan2025wan}, which produces 5s clips at 16 FPS and $832\times480$ resolution.
We first adapt the pretrained model into a few-step causal-attention model using a causal-forcing~\citep{causalforcing} DMD pipeline on VidProM~\citep{vidprom} data, while enabling our \memoryquery{} and Unified \relativerope{} Recipe. 
We then perform streaming long tuning on a 60s sequence that contains a single prompt switch~\citep{longlive}. 
The memory encoder consists of $L_{\mathrm{enc}}{=}2$ cross-attention layers with hidden dimension $1536$, $12$ heads, and head dim $128$.
We instantiate $T_{\Qset}{=}N_{\Qset} (3)$ memory frames $\times$ $S (1560)$ tokens per frame $=4680$ query tokens.
The local KV window stores $N_W{=}9$ frames, the sink consists of $N_S{=}3$ frames, and the pretrained DiT's maximum temporal RoPE id is $\fmax{=}20$. More details are in  \S\ref{appendix:implementation}.

% \vspace{-1pt}
\subsection{Long Video Generation}
\label{sec:exp-long}
% \vspace{-1pt}

\textbf{Protocol.}
We evaluate \ourmethod{} on $30$s and $240$s long-video generation following prior work~\citep{infinityrope,longlive}.
For $30$s, we use the official VBench-Long prompt set~\citep{huang2025vbench++} and aggregate its $16$ metrics with official weights into quality and semantic scores.
We also conduct a forced-choice user study, detailed in \S\ref{appendix:user_study}: $30$ users evaluate $50$ randomly sampled results with anonymous, randomized model order, and select the best video by overall quality and semantic~(caption) alignment.
We report each method's selection rate as the user preference.
For ultra-long generation, following prior work~\citep{infinityrope,deepforcing}, we evaluate $240$s videos on $128$ randomly sampled MovieGen prompts~\citep{moviegen}, and use aesthetic quality, background consistency, dynamic degree, imaging quality, motion smoothness, and subject consistency, aggregated into the quality score with official VBench-Long weights~\citep{huang2025vbench++}.
We also report user preference for semantic coherence and overall quality.

\begin{figure}[t]
  \centering
  % \vspace{-20pt}
  \includegraphics[width=\linewidth]{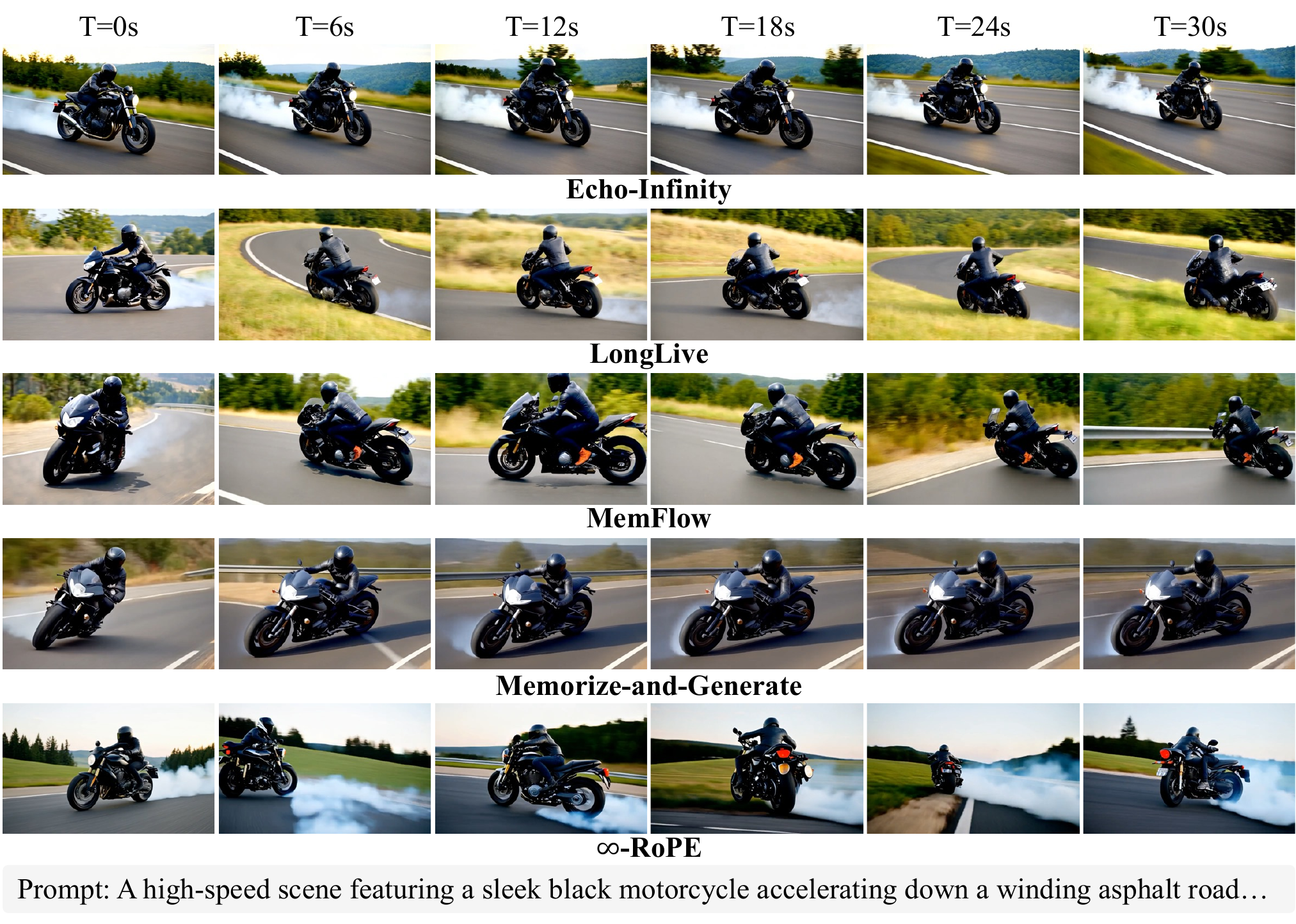}
  \vspace{-10pt}
  \caption{\textbf{$\textbf{30}\,$s Long Video Generation Result Comparison.} More visualizations are in the supplementary.}
  % \vspace{-7pt}
  \label{fig:exp-vis-30s}
\end{figure}

\textbf{Results.} We mainly compare with recent DMD-based long and real-time autoregressive video diffusion methods~\citep{improved-dmd,dmd}, which are generally stronger than conventional autoregressive baselines such as SkyReels-V2~\citep{skyreelsv2} and MAGI-1~\citep{magi1}: LongLive~\citep{longlive}, MemFlow~\citep{memflow}, Memorize-and-Generate~\citep{memorizeandgenerate}, and $\infty$-RoPE~\citep{infinityrope}.
As shown in Tab.~\ref{tab:long-merged}, \ourmethod{} performs best at both horizons with real-time throughput~(18.5 FPS). It notably improves $30$s user preference~($\textbf{59.53}$ vs. $\textbf{14.73}$), and the gains further grow at $240$s, improving quality from $\textbf{79.99}$ to $\textbf{81.23}$ and user preference from $\textbf{14.13}$ to $\textbf{71.67}$.
Figs.~\ref{fig:exp-vis-240s},~\ref{fig:exp-vis-30s} show our better long-range identity and scene coherence. As in Fig.~\ref{fig:exp-vis-240s}, LongLive and MemFlow suffer from identity drift, with missing face masks, changing astronaut counts, and varying outfits. 
Memorize-and-Generate shows severe degradation at long horizons, likely due to accumulated information loss under fixed-ratio history compression, while $\infty$-RoPE shows over-exposure and color shifts under unsolved RoPE train-test mismatch.

\begin{figure}[t]
  \centering
  % \vspace{-22pt}
  \includegraphics[width=\linewidth]{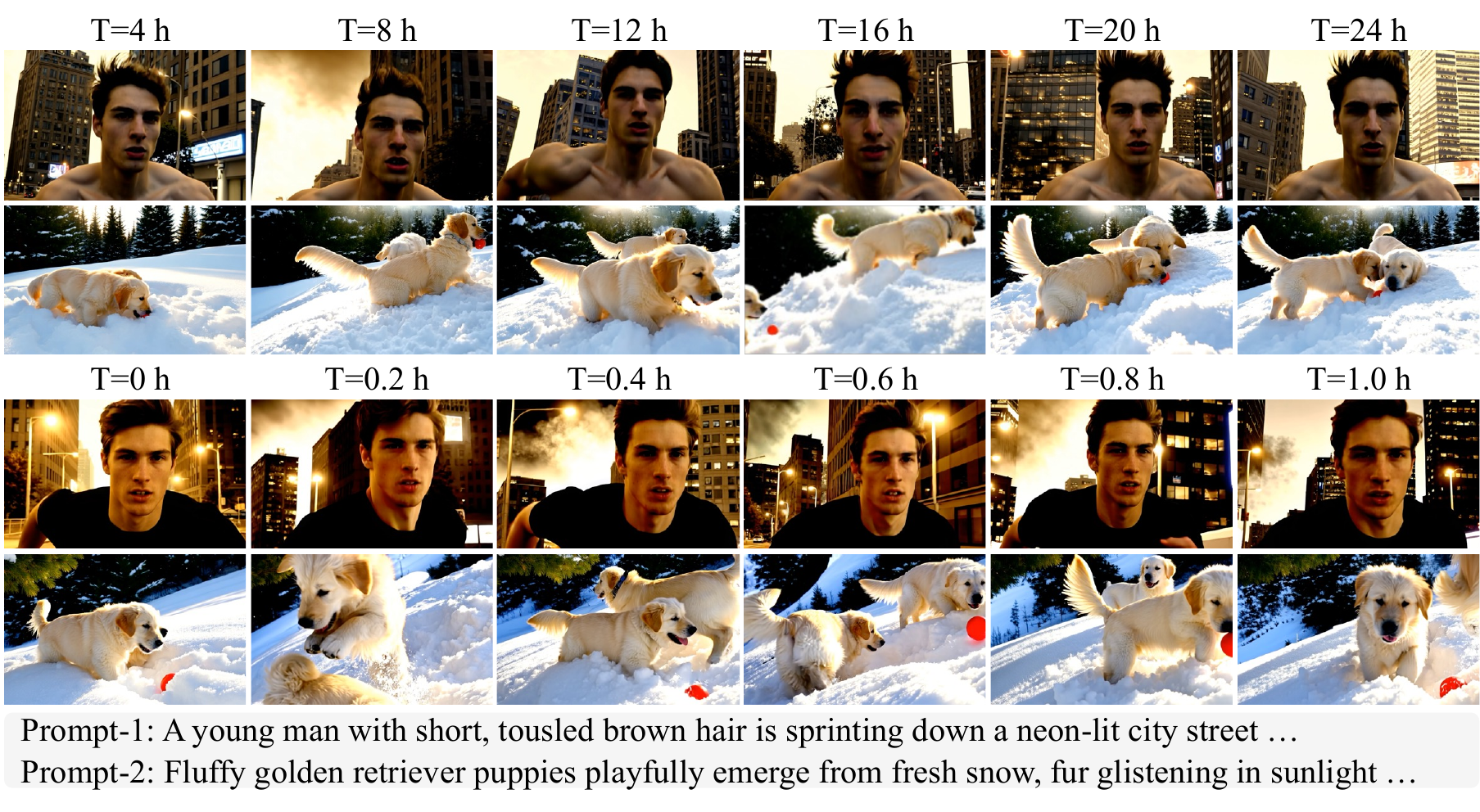}
  % \vspace{-20pt}
  \caption{\textbf{$\textbf{1}$-Hour and \textbf{24}-Hour Video Generation Result Comparison.}}
  % \vspace{-5pt}
  \label{fig:exp-vis-hour}
\end{figure}

\textbf{Infinite Video Generation.}
With evolving learnable memory queries and a unified relative RoPE recipe that removes train-inference RoPE mismatch and fixed positional limits, \ourmethod{} naturally supports infinite video generation.
We verify this by sampling $15$ MovieGen prompts~\citep{moviegen} and running real-time inference for $1$ and $24$ hours, as shown in Fig.~\ref{fig:exp-vis-hour}.
We compare the results generated from the same prompts at the two lengths.
The results show that \ourmethod{} preserves stable visual quality and strong consistency from $\textbf{1}$ to $\textbf{24}$ hours.
\textbf{We strongly encourage viewing \href{https://yxbian23.github.io/Echo-Infinity/}{our
 project page} for qualitative visualization}, since current quantitative protocols~\citep{huang2025vbench++} are too compute-heavy to evaluate such extremely long videos.

\begin{table}[!t]
    \centering
    \caption{\textbf{Single-Prompt 5s Video Evaluation} on VBench's official benchmark~\citep{huang2023vbench}.}
    % \vspace{-5pt}
    \label{tab:short-5s}
    \small
    \setlength{\tabcolsep}{4pt}
    \renewcommand{\arraystretch}{0.82}
    \setlength{\aboverulesep}{1pt}
    \setlength{\belowrulesep}{1pt}
    \setlength{\extrarowheight}{1pt}
    \resizebox{\linewidth}{!}{%
    \begin{tabular}{lccccc}
    \toprule
    \multirow{2}{*}{\textbf{Model}} & \multirow{2}{*}{\textbf{\#Params}} & \multirow{2}{*}{\textbf{Throughput (FPS) $\uparrow$}} & \multicolumn{3}{c}{\textbf{Evaluation scores $\uparrow$}} \\
    \cmidrule(lr){4-6}
    & & & \textbf{Total} & \textbf{Quality} & \textbf{Semantic} \\
    \midrule
    \multicolumn{6}{l}{\textit{Diffusion Models}} \\
    LTX-Video~\citep{ltxvideo} & 1.9B & 8.98 & 80.00 & 82.30 & 70.79 \\
    Wan-2.1~\citep{wan2025wan} & 1.3B & 0.78 & 84.26 & 85.30 & 80.09 \\
    \midrule
    \multicolumn{6}{l}{\textit{Autoregressive Models}} \\
    SkyReels-V2~\citep{skyreelsv2}               & 1.3B & 0.49 & 82.67 & 84.70 & 74.53 \\
    MAGI-1~\citep{magi1}                         & 4.5B & 0.19 & 79.18 & 82.04 & 67.74 \\
    Self Forcing, chunk-wise~\citep{selfforcing} & 1.3B & 17.0 & 83.08 & 83.97 & 79.53 \\
    Causal Forcing, chunk-wise~\citep{causalforcing} & 1.3B & 17.0 & 83.94 & 84.59 & \underline{81.35} \\
    \midrule
    \multicolumn{6}{l}{\textit{Long Autoregressive Models}} \\
    LongLive~\citep{longlive}                         & 1.3B & \underline{20.7} & 83.29 & 84.09 & 80.06 \\
    MemFlow~\citep{memflow}                           & 1.3B & 18.7 & 83.62 & 84.52 & 80.02 \\
    Memorize-and-Generate~\citep{memorizeandgenerate} & 1.3B & \textbf{21.7} & 84.06 & 84.84 & 80.96 \\
    \midrule
    % \rowcolor{yellow!20}
    \textbf{\ourmethod{} (w/o Memory Update)} & 1.3B & 18.9 & \underline{84.57} & \underline{85.51} & 80.80 \\
    \textbf{\ourmethod{} (w/ Memory Update)} & 1.3B & 18.5 & \textbf{85.35} & \textbf{86.32} & \textbf{81.49} \\
    \bottomrule
    \end{tabular}}
    % \vspace{-12pt}
\end{table}

\subsection{Interactive Long Video Generation}
\label{sec:exp-interactive}

Our \memoryquery{} and unified relative RoPE recipe improve long-range quality and consistency, orthogonal to interactive generation optimization. Directly applied to multi-prompt $60$ s interactive generation, they still improve long-horizon quality, as shown in Tab.~\ref{tab:interactive-60s} and Fig.~\ref{fig:exp-vis-interactive}.
We follow MemFlow's protocol~\citep{memflow} and evaluate on $100$ narrative scripts, each with six successive 10-second prompts.
We use VBench-Long metrics~\citep{huang2025vbench++} for visual quality and CLIP scores~\citep{hessel2021clipscore} for clip-wise semantic adherence at 10-second intervals.
Tab.~\ref{tab:interactive-60s} and Fig.~\ref{fig:exp-vis-interactive} show that \ourmethod{} achieves the best overall quality and competitive prompt alignment.

% \vspace{-3pt}
\subsection{Short Video Generation}
\label{sec:exp-short}
% \vspace{-3pt}

\begin{figure}[t]
  \centering
  % \vspace{-22pt}
  \includegraphics[width=\linewidth]{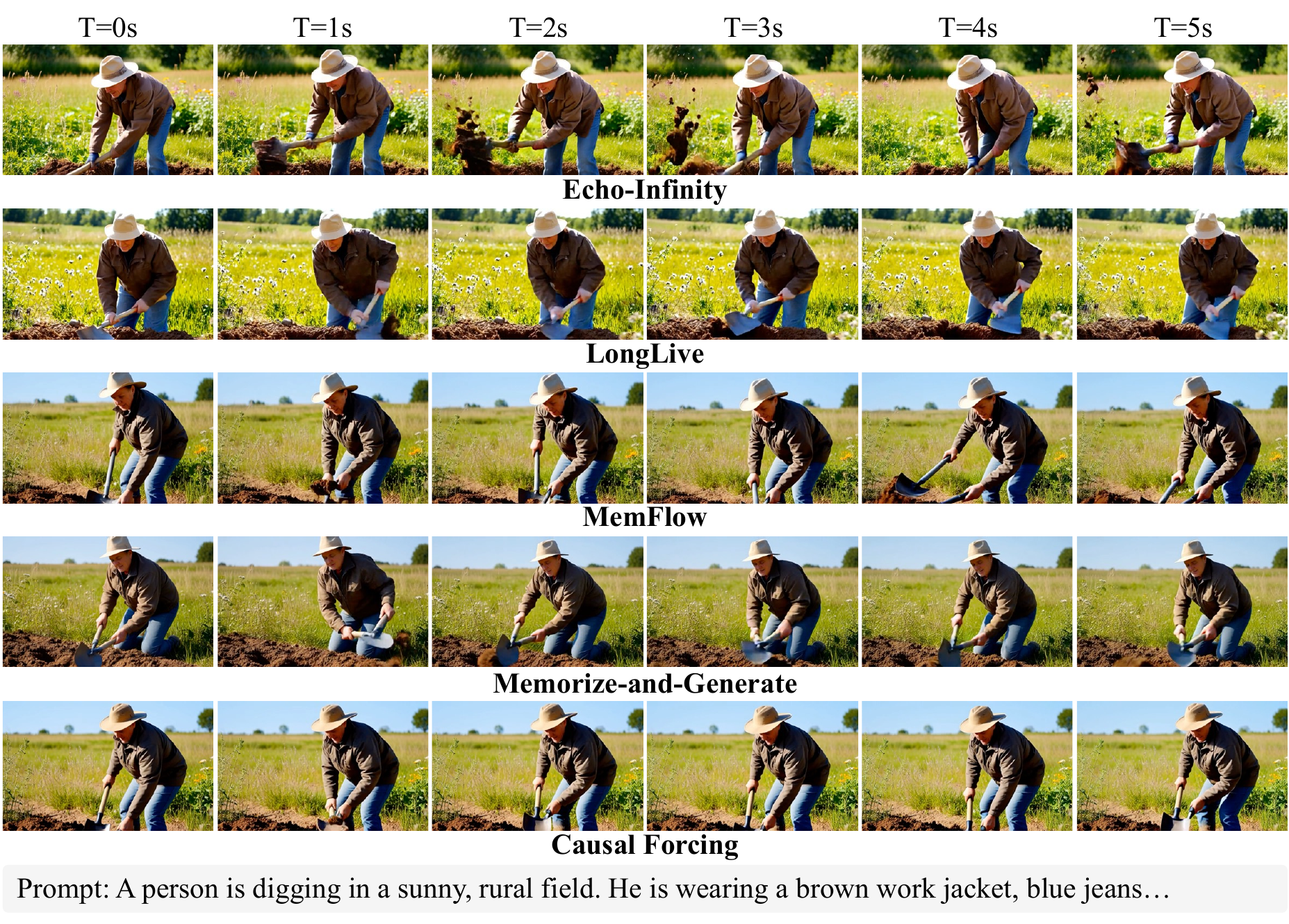}
  % \vspace{-15pt}
  \caption{\textbf{$\textbf{5}\,$s Short Video Generation Result Comparison.} More visualizations are in the supplementary.}
  % \vspace{-5pt}
  \label{fig:exp-vis-5s}
\end{figure}

We further evaluate \ourmethod{} on standard $5$s video generation using official VBench prompts~\citep{huang2023vbench}, comparing with open-source models of comparable scale~\citep{wan2025wan,skyreelsv2,selfforcing,causalforcing,longlive,memflow,memorizeandgenerate}.
As shown in Tab.~\ref{tab:short-5s}, \ourmethod{} without memory updates already outperforms all baselines with a total score of $84.57$, showing that the optimized initial $\Qset$ itself serves as an effective generation prior.
Enabling memory updates further improves the total, quality, and semantic scores to $85.35$, $86.32$, and $81.49$, respectively, by allowing the memory to adapt dynamically.
As in the digging example of Fig.~\ref{fig:exp-vis-5s}, \ourmethod{} generates plausible soil splashes after shovel motions, whereas other methods show weaker motion effects or less realistic interactions.

% \vspace{-3pt}
\subsection{Ablation Studies}
\label{sec:exp-ablation}
% \vspace{-3pt}

\begin{table}[t]
    \centering
    \caption{\textbf{Ablation Studies.} We ablate \ourmethod{} on $240$s generation of $128$ MovieGen prompts. More ablation results and analysis are in \S\ref{appendix:ablation} due to page limit.}
    % \vspace{-4pt}
    \label{tab:ablation}
    \small
    \setlength{\tabcolsep}{2pt}
    \resizebox{\linewidth}{!}{%
    \begin{tabular}{lccccc}
    \toprule
    \textbf{Model} & \textbf{Subject Consistency} $\uparrow$ & \textbf{Background Consistency} $\uparrow$ & \textbf{Dynamic Degree} $\uparrow$ & \textbf{Aesthetic Quality} $\uparrow$ & \textbf{CLIP Score} $\uparrow$ \\
    \midrule
    (a) w/o \memoryquery{}                                   & 96.15 & 95.27 & 64.78 & 58.60 & 32.78 \\
    (b) w/o Unified \relativerope{} Recipe                                  & 96.34 & 95.81 & 64.05 & 59.83 & 33.12 \\
    (c) w/ Self Forcing ODE init                             & 96.85 & 96.12 & 52.04 & 58.49 & 34.07 \\
    \midrule
    % \rowcolor{yellow!20}
    \textbf{\ourmethod{} (Ours)}    & 96.58 & 95.93 & 68.61 & 58.67 & 34.19 \\
    \bottomrule
    \end{tabular}}
    % \vspace{-15pt}
\end{table}

We ablate our core designs on $240$s video generation using $128$ randomly sampled prompts from MovieGen~\citep{moviegen}.
As shown in Tab.~\ref{tab:ablation}, each design contributes to the final \ourmethod{}:
\textbf{(a) \memoryquery{}.} Removing memory queries substantially degrades consistency and dynamic degree, due to lost historical information.
\textbf{(b) Unified \relativerope{} Recipe.} Replacing relative RoPE with absolute RoPE degrades most quality and consistency metrics, as long video inference leads to unseen positional indices and causes out-of-distribution failures.
\textbf{(c) ODE Initialization.} We employ Causal Forcing~\citep{causalforcing}, which uses a refined ODE over Self Forcing~\citep{selfforcing}. Switching to Self Forcing yields similar performance, except for the expected dynamic degree drop due to its ODE limitation~\citep{causalforcing}, demonstrating robustness to ODE initialization.
% 

% \vspace{-5pt}
\section{Conclusion}
\label{sec:conclusion}

We presented \ourmethod{}, an end-to-end memory-learning framework towards real-time infinite video generation.
Instead of handcrafted KV-cache schedules or fixed-ratio compression, \ourmethod{} learns compact evolving \memoryquery{} that filter, abstract, and compress arbitrary-length history at constant cost.
Together with our Unified \relativerope{} Recipe, which keeps temporal RoPE indices within the trained range in both training and inference, \ourmethod{} removes two key bottlenecks of infinite video generation: unbounded memory growth and positional extrapolation.
Extensive experiments show that \ourmethod{} achieves state-of-the-art performance on long, short, and interactive video generation, while showing promising results for over $24$ hours and $1.3M$ frames in real time.
We hope this work opens up possibilities for future infinite video generation with persistent memory, controllable long-range dynamics, and unbounded horizons.

\noindent \textbf{Limitations and Future Works.}
% \paragraph{Limitations.} 
\ourmethod{} has several limitations.
(1) The memory queries and unified relative RoPE recipe are designed to improve the quality and consistency for long video generation. Optimizing them for interactive generation remains open.
(2) Limited by the base model scale and generative capacity, \ourmethod{} may show reduced stability for hour-scale or longer videos with dynamic scenes.
For future work, we believe:
(1) Studying the semantics of memory queries could enable controllable retrieval-augmented video generation.
(2) Distilling \ourmethod{} into a one-step model could improve real-time throughput.
% 
% More details are in \S\ref{appendix:limitations}.
% \input{secs/author_list}

\clearpage
\bibliographystyle{plainnat}
\bibliography{references}

% \clearpage
% \beginappendix
\newpage

\appendix

In this appendix, we first provide additional qualitative visualizations of \ourmethod{} on $30\,$s, $240\,$s, and $60\,$s interactive long video generation~(\S\ref{appendix:visualizations}).
We then detail the implementation of \ourmethod{}, including the memory-encoder architecture, optimization recipe, and training schedule~(\S\ref{appendix:implementation}), and review related work on long video generation, memory mechanisms, and rotary positional embeddings~(\S\ref{appendix:related_works}).
Building on the experiments in the main paper, we report extended ablations on each component of \ourmethod{}~(\S\ref{appendix:ablation}) and describe the protocol of our forced-choice user study~(\S\ref{appendix:user_study}).
We also include additional discussion of our limitations and future work~(\S\ref{appendix:limitations}).
Finally, we discuss the broader societal impacts~(\S\ref{appendix:broader_impacts}).

\section{More Visualizations}
\label{appendix:visualizations}

\begin{figure}[t]
  \centering
  % \vspace{-20pt}
  \includegraphics[width=\linewidth]{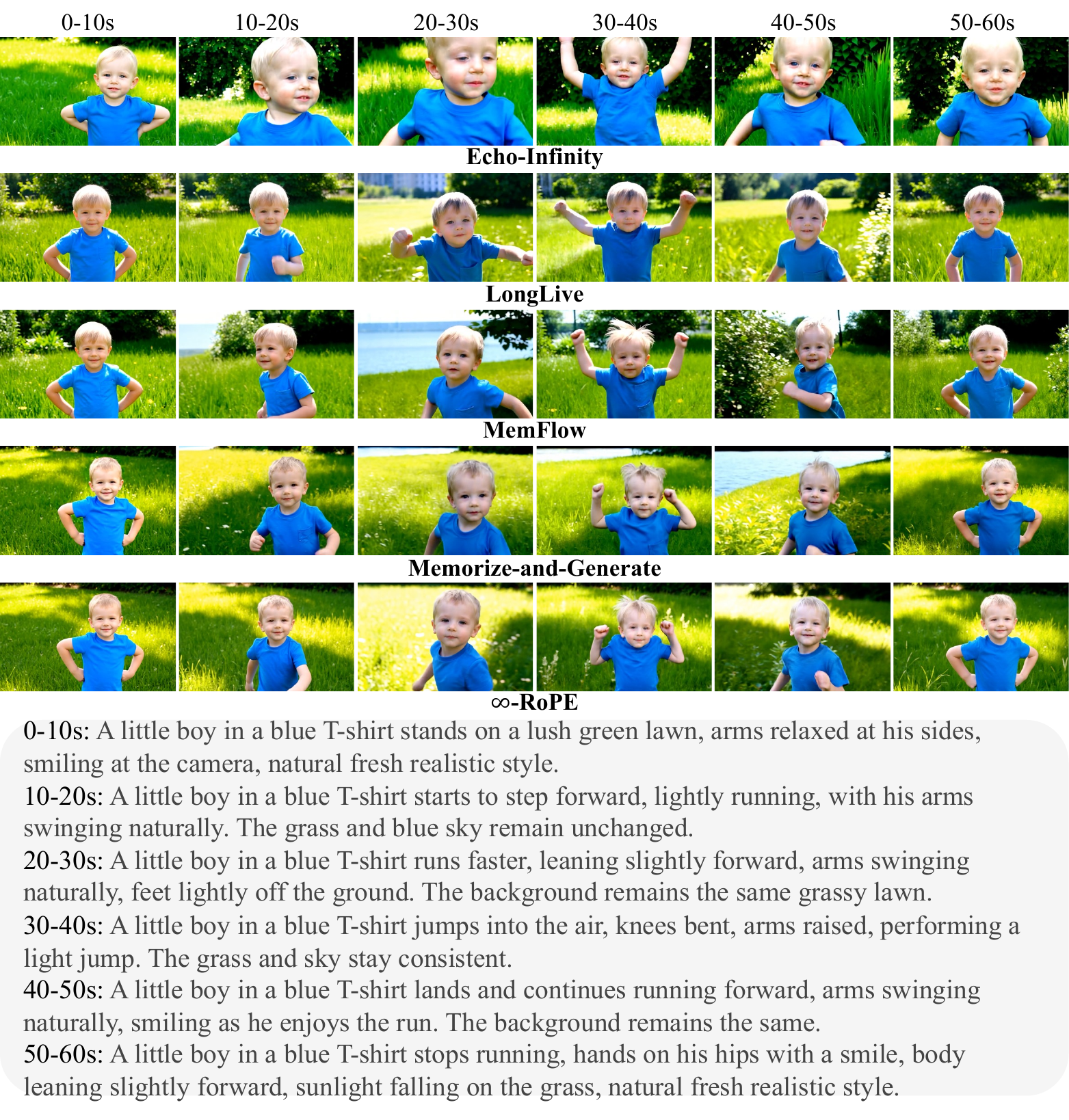}
  \vspace{-10pt}
  \caption{\textbf{$\textbf{60}\,$s Interactive Long Video Generation Comparison.}}
  % \vspace{-15pt}
  \label{fig:exp-vis-interactive}
\end{figure}

\begin{figure}[t]
  \centering
  % \vspace{-20pt}
  \includegraphics[width=\linewidth]{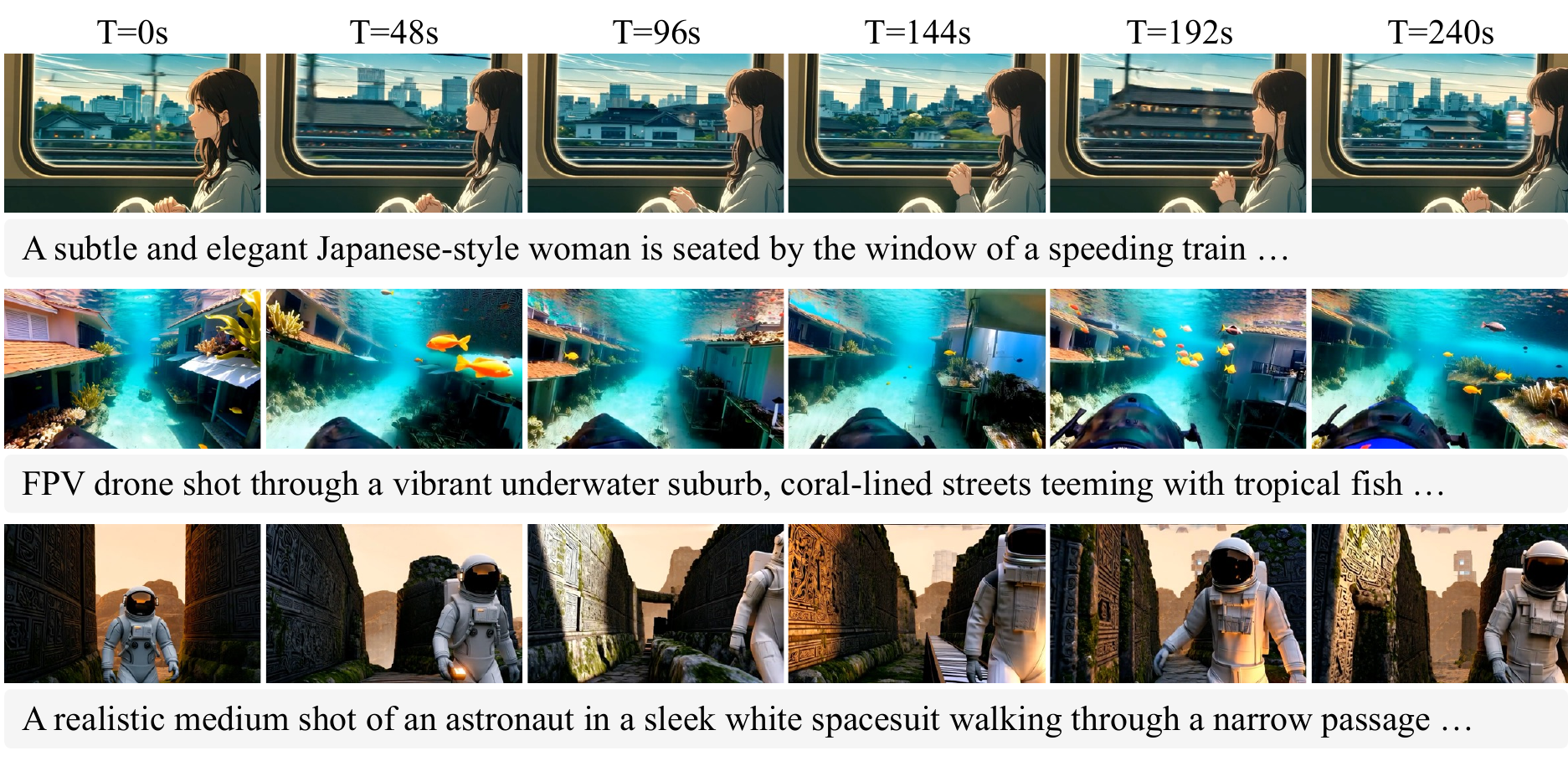}
  \vspace{-10pt}
  \caption{\textbf{$\textbf{240}\,$s Long Video Generation Visualization.}}
  % \vspace{-15pt}
  \label{fig:appendix-vis-240s}
\end{figure}

\begin{figure}[t]
  \centering
  % \vspace{-20pt}
  \includegraphics[width=\linewidth]{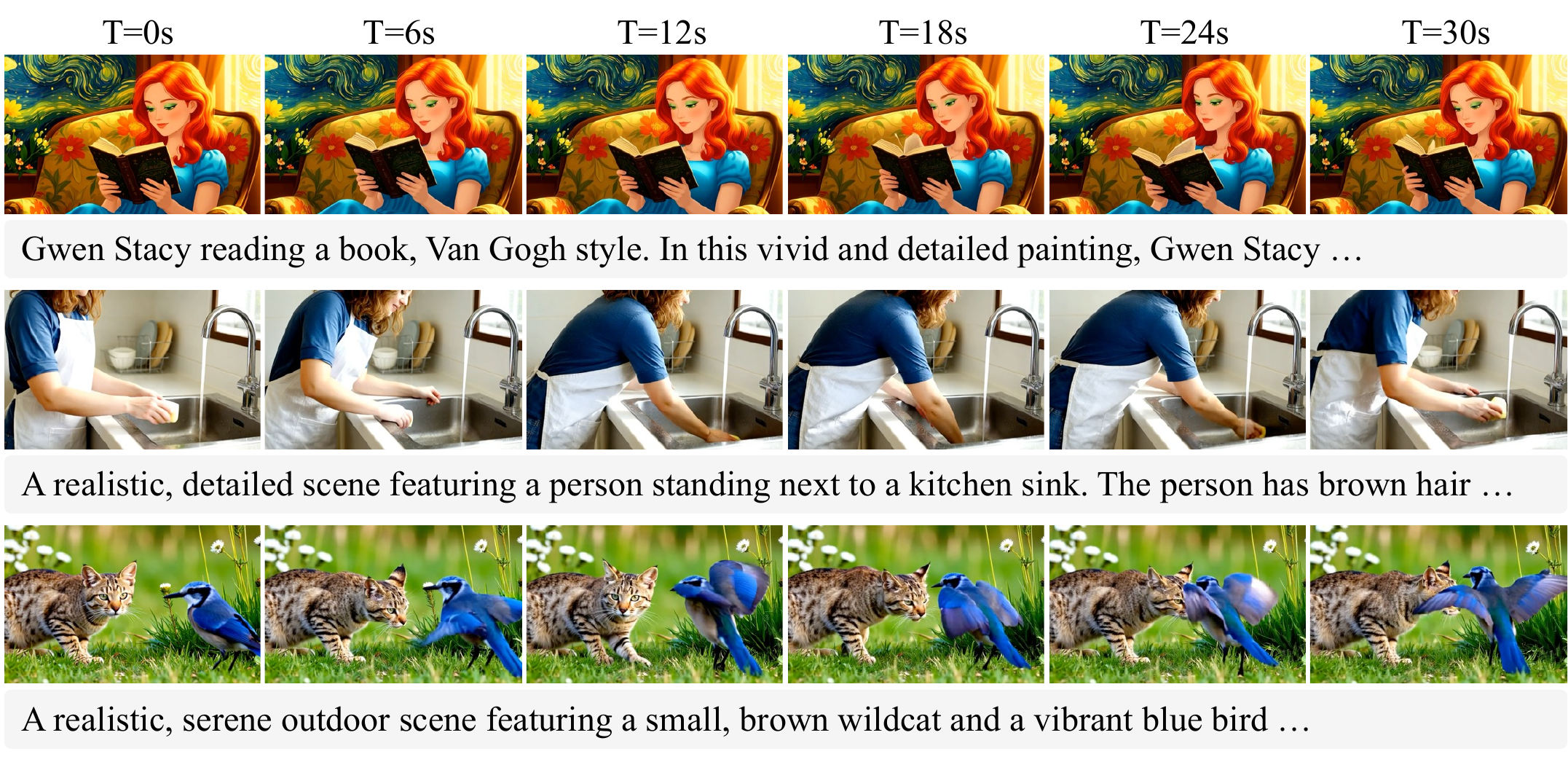}
  \vspace{-10pt}
  \caption{\textbf{$\textbf{30}\,$s Long Video Generation Visualization.}}
  % \vspace{-15pt}
  \label{fig:appendix-vis-30s}
\end{figure}

To further demonstrate \ourmethod{}'s performance, we provide more visualization cases (long interactive video generation, and more 240s / 30s video generation) in Fig.~\ref{fig:exp-vis-interactive}, Fig.~\ref{fig:appendix-vis-240s}, and Fig.~\ref{fig:appendix-vis-30s}. 
\textbf{We strongly encourage viewing \href{https://yxbian23.github.io/Echo-Infinity/}{our project page} for qualitative visualization.}

\section{Implementation Details}
\label{appendix:implementation}
We implement \ourmethod{} on Wan2.1-T2V-1.3B~\citep{wan2025wan}, which produces 5s clips at 16 FPS and $832\times480$ resolution.
We first adapt the pretrained model into a few-step causal-attention model using a causal-forcing~\citep{causalforcing} DMD pipeline on VidProM~\citep{vidprom} data, while enabling our \memoryquery{} and Unified \relativerope{} Recipe. 
Notably, all KV caches are stored in their pre-RoPE states, allowing the subsequent unified relative RoPE recipe to rotate them directly to the desired RoPE IDs when they are used in the DiT causal attention or memory queries update.
We then perform streaming long tuning on a 60s sequence that contains a single prompt switch following LongLive's protocol~\citep{longlive}. 
By default, we use chunk size $B=3$, following the same setting as previous works~\citep{longlive,memflow,infinityrope,memorizeandgenerate}.
The memory encoder consists of $L_{\mathrm{enc}}{=}2$ cross-attention layers with hidden dimension $1536$, $12$ heads, and head dim $128$.
The gated residual is initialized with bias $2.0$, so $\mathbf{g}\!\approx\!0.88$ at training step $0$.
We instantiate $T_{\Qset}{=}N_{\Qset} (3)$ memory frames $\times$ $S (1560)$ tokens per frame $=4680$ query tokens.
The local KV window stores $N_W{=}9$ frames, the sink consists of $N_S{=}3$ frames, and the pretrained DiT's maximum temporal RoPE id is $\fmax{=}20$.
Following LongLive~\citep{longlive}, optimization uses AdamW for both fake generator and student model with learning rates \(\mathrm{lr}_{\mathrm{fake}}=1.0\times10^{-5}\) and \(\mathrm{lr}_{\mathrm{student}}=2.0\times10^{-6}\); we set \(\beta_{1,\mathrm{fake}}=0.0\), \(\beta_{2,\mathrm{fake}}=0.999\) for the fake generator and \(\beta_{1,\mathrm{student}}=0.0\), \(\beta_{2,\mathrm{student}}=0.999\) for the student. Training is conducted on \(64\) GPUs with one sample per GPU (global batch size \(=64\)). We apply EMA to the student with decay \(0.99\), starting at step \(200\). For the \(60\,\mathrm{s}\) setting, we train for \(3{,}000\) iterations.
Memory-encoder parameters use a learning rate of $5\times$ the generator's.
For interactive video generation, which is not the focus of this work, we inherit the recache operation from LongLive~\citep{longlive}. For our added memory queries, at each recache step, we update them using all local KV caches after recaching, yielding an analogous recache operation for the evolving memory queries and moving it toward the new prompt.

\paragraph{Compute.}
All training runs are conducted on $64$ NVIDIA H200 GPUs with one sample per GPU. The first-stage causal-forcing DMD adaptation is trained for $400$ iterations and takes about $3$ hours. The second-stage streaming long tuning is trained for $3{,}000$ iterations and takes about $12$ hours. The main training run therefore uses about $960$ H200 GPU-hours. Unless otherwise specified, throughput is measured on a NVIDIA H100.

\paragraph{Evaluation reproducibility.}
All benchmarks follow the standard protocols used in prior long-video generation and VBench/VBench-Long evaluations~\citep{huang2023vbench,huang2025vbench++,longlive,infinityrope}.
For all automatic evaluations, each prompt is generated with five random seeds, and we report the average score over seeds and prompts.
For all baselines, we run with the same Wan2.1-T2V-1.3B backbone and their officially public checkpoints, same resolution, same number of sampling steps, same prompt sets, and the same hardware.

\section{Related Works}
\label{appendix:related_works}
\subsection{Long Video Generation}

Modern video diffusion models~\citep{cogvideox,hunyuan-video,wan2025wan,ltxvideo,moviegen,skyreelsv2} mostly adopt bidirectional Diffusion Transformers~(DiTs) that jointly denoise an entire short clip and have obtained rapid progress in various downstream applications~\citep{bai2025recammaster,bian2025video,bian2025videopainter,bian2025motioncraft}.
Their quadratic attention cost, together with the need to denoise the full clip before any frame can be emitted, makes this paradigm unsuitable for streaming long-video generation.
Recent work turns to autoregressive paradigms~\citep{nova,pyramidflow,magi1} similar to large language models' inference~\citep{yang2025qwen3,comanici2025gemini,cai2024medusa} that generate video chunk by chunk under causal attention.
Building on Diffusion Forcing~\citep{diffusion-forcing}, which uses staircase streaming denoising, and Self-Forcing~\citep{causvid,selfforcing}, which builds on DMD distillation~\citep{improved-dmd,dmd}, these methods push video generation toward minute-scale duration and real-time streaming~\citep{longlive,rollingforcing,selfforcingpp,framepack,infinityrope}.
Despite this progress, unbounded KV-cache growth and temporal RoPE train-test mismatch remain open challenges.

\subsection{Memory Mechanisms in Long Video Generation}

Current methods mainly address the unbounded KV cache based on specially designed memory mechanisms:
\noindent\textbf{(a) Window Truncation.}\quad
These methods retain a bounded local window plus a few \sinkframe{} and discard others, leading to inevitable history loss~\citep{longlive,rollingforcing}.
\noindent\textbf{(b) Hand-Crafted KV-cache Management.}\quad
A second family augments the local window with rule-based scheduling that decides which evicted KVs to keep~\citep{memflow,contextforcing,relaxforcing,packforcing,anchorforcing,groundedforcing}.
For example, MemFlow~\citep{memflow} curates a memory bank by textual retrieval.
However, these rules are tuned offline and remain bound by the given window length.
\noindent\textbf{(c) Heuristic Compression.}\quad
Rather than selecting evicted KVs, a third family replaces the evicted history with compressed representations~\citep{memorizeandgenerate,lovic,framepack,pretrainingframepreservation}.
Memorize-and-Generate~\citep{memorizeandgenerate} decouples memory compression and frame generation by compressing historical information into compact KVs via reconstruction, which are consumed by a separate generator.
LoViC~\citep{lovic} compresses long video-text context into latent context tokens with a separately trained autoencoder and feeds them to a DiT decoder.
These methods improve the usable temporal context, but their compressed states are still tied to prescribed compression ratios, compression schedules, or separate reconstruction/compression stages.
In contrast, \ourmethod{} recurrently consolidates evicted causal KVs into a persistent memory query state used directly as a plug-in KV source, learning what history to preserve end-to-end under the long-video generation objective while keeping the active memory footprint independent of sequence length.

\paragraph{Failure mode of fixed-ratio compression.}
Let $T$ denote the number of generated frames and $C$ denote a fixed compression ratio.
If one compressed unit is retained for every $C$ historical units, the compressed history still contains about $\lceil T/C \rceil$ units, giving a memory footprint of
\[
\mathcal{O}(T/C).
\]
Thus, even after compression, the memory footprint remains unbounded as $T \to \infty$.
To keep the memory bounded by a constant budget $M$, the compression ratio would have to grow with the horizon, i.e., $C=\Omega(T/M)$, which changes the operating point over time and becomes increasingly aggressive.
\ourmethod{} avoids this fixed-ratio failure mode by adopting a fixed-capacity view: all evicted history is recurrently consolidated into $\Qset$ with $T_{\Qset}=N_Q S$ tokens.
Once the sink size $N_S$, local window size $N_W$, chunk size $B$, and memory-query budget $T_{\Qset}$ are fixed, the active context size is
\[
\mathcal{O}(N_S S + N_W S + T_{\Qset} + B S)=\mathcal{O}(1)
\]
with respect to the generated sequence length $T$.
Therefore, the effective compression ratio can grow with video length, while the algorithm and memory budget remain unchanged.

VideoSSM~\citep{videossm} instead compresses long-range history with a state-space global memory while retaining a local context window for recent motion and details.
However, VideoSSM realizes this state through architecture-level SSM modules, whereas \ourmethod{} keeps the DiT attention blocks unchanged and injects layer-shared \memoryquery{} as a plug-in KV source.
This makes our memory mechanism compatible with any pretrained attention backbone and our unified \relativerope{} recipe, while still providing fixed-size recurrent memory for arbitrary-length history.

\subsection{Rotary Positional Embedding for Long Video Generation}

% \noindent\textbf{3D-RoPE for video DiTs.}\quad
Modern video DiTs apply 3D rotary positional embeddings (RoPE) independently along the temporal, height, and width axes~\citep{rope}.
However, under autoregressive rollouts, the temporal index grows beyond the bounded range seen in pretraining (e.g., a maximum of 20 for Wan-2.1~\citep{wan2025wan}), leading to rapid quality collapse and eventual overflow.
%
% \noindent\textbf{Inference-based mitigations.}\quad
Some concurrent training-free methods on frozen backbones~\citep{infinityrope,deepforcing,memrope} observe these challenges but only mitigate the overflow risk at inference time by adopting relative RoPE, leaving the train-test inconsistency unsolved.
These test-time patches do not change the index distribution the backbone saw during training, so the gap persists in the weights.
To our knowledge, \ourmethod{} is the first to eliminate the temporal RoPE mismatch during both training and inference, using a unified relative RoPE schedule that keeps every active temporal index within the same range in all stages.

\section{Ablation Studies}
\label{appendix:ablation}

We ablate our core designs on $240$s video generation using $128$ randomly sampled prompts from MovieGen~\citep{moviegen}. Results are summarized in Tab.~\ref{tab:ablation-full}.

\begin{table}[t]
    \centering
    \caption{\textbf{Ablation Studies.} We ablate \ourmethod{} on $240$s generation of $128$ MovieGen prompts.}
    \label{tab:ablation-full}
    \small
    \setlength{\tabcolsep}{2pt}
    \resizebox{\linewidth}{!}{%
    \begin{tabular}{lccccc}
    \toprule
    \textbf{Model} & \textbf{Subject Consistency} $\uparrow$ & \textbf{Background Consistency} $\uparrow$ & \textbf{Dynamic Degree} $\uparrow$ & \textbf{Aesthetic Quality} $\uparrow$ & \textbf{CLIP Score} $\uparrow$ \\
    \midrule
    (a) w/o \memoryquery{}                                   & 96.15 & 95.27 & 64.78 & 58.60 & 32.78 \\
    (b) w/o Unified \relativerope{} Recipe                                  & 96.34 & 95.81 & 64.05 & 59.83 & 33.12 \\
    (c) w/ Self Forcing ODE init                             & 96.85 & 96.12 & 52.04 & 58.49 & 34.07 \\
    (d) \memoryquery{} number $N_{\Qset}{=}1$                    & 94.72 & 94.03 & 67.39 & 57.84 & 33.62 \\
    (e) \memoryquery{} number $N_{\Qset}{=}5$                    & 96.44 & 96.08 & 69.47 & 58.81 & 34.05 \\
    (f) w/o Gate                                             & 95.31 & 94.62 & 68.43 & 58.49 & 34.02 \\
    \midrule
    % \rowcolor{yellow!20} 
    \textbf{\ourmethod{} (Ours)}    & 96.58 & 95.93 & 68.61 & 58.67 & 34.19 \\
    \bottomrule
    \end{tabular}}
\end{table}

As shown in Tab.~\ref{tab:ablation-full}, each design contributes to the final \ourmethod{}:
\textbf{(a) \memoryquery{}.} Removing memory queries substantially degrades consistency and dynamic degree, as historical information is inevitably lost.
\textbf{(b) Unified \relativerope{} Recipe.} Replacing relative RoPE with the original absolute RoPE degrades most quality and consistency metrics, since absolute RoPE requires unseen positional indices at inference time and causes severe out-of-distribution generalization failure.
\textbf{(c) ODE Initialization.} We build on Causal Forcing~\citep{causalforcing}, which uses a refined ODE solver over Self Forcing~\citep{selfforcing}. Switching to Self Forcing yields similar performance, except for the expected dynamic-degree drop due to its ODE limitation~\citep{causalforcing}, demonstrating robustness to ODE initialization.
\textbf{(d) Memory Query Number.} Our default setting uses $N_{\Qset}=3$ memory-query frames. 
Using only $1$ frame performs much worse, likely because too few memory queries cannot capture sufficient long-term history, leading to incomplete, unstable memory that confuses the current generation under the same training budget. 
Increasing to $5$ frames gives only marginal gains while reducing inference speed~(from $18.5$ FPS to $17.8$ FPS).
\textbf{(e) Gate Mechanism.} The memory-query update in Eq.~\ref{eq:memq-update}~(\S\ref{sec:method-memquery}) uses a sigmoid gate to interpolate between the previous memory state $\Qset$ and the newly encoded $\widetilde{\Qset}$. Replacing it with a residual update, i.e., $\Qset \leftarrow \Qset + \widetilde{\Qset}$, noticeably reduces subject and background consistency, while dynamic degree, aesthetic quality, and CLIP score remain nearly stable.
Without the gate, each eviction injects $\widetilde{\Qset}$ into $\Qset$ without selectively preserving history. This weakens memory as a persistent history representation and degrades long-term consistency.

\section{User Study}
\label{appendix:user_study}

We conduct a forced-choice user study to evaluate human preference for 30s/240s long-video generation. For each evaluated duration, we randomly sample $50$ prompts from the corresponding benchmark. For each prompt, we collect one generated video from each method under the same prompt and duration setting, forming one result group.

We recruit $30$ video generation researchers to evaluate all result groups. In each trial, participants are shown the generated videos from all methods in anonymized and randomized order, without access to model names. Participants are asked to select the single best video by jointly considering \emph{overall visual quality}, \emph{temporal consistency}, \emph{motion naturalness}, and \emph{caption alignment}. The display order is randomized independently for each trial and each participant to reduce positional bias.
For 30s videos, participants are required to watch each video in full before making a selection. For 240s videos, to keep the evaluation workload feasible while still exposing participants to both early and late rollout behavior, participants are required to watch at least $60$ seconds in total from each video. Specifically, the first $10$ seconds and the final $10$ seconds, i.e., $230$--$240$s, must be watched in full. For the middle segment, i.e., $10$--$230$s, participants are allowed to freely control the progress bar and inspect segments of interest, but are required to watch at least $40$ additional seconds before making a selection. This protocol ensures that participants evaluate both the initial visual quality and the end-of-rollout consistency, while keeping the total annotation time practical for 240s videos.

For each method, we compute its user preference rate as the percentage of times its result is selected:
\[
\mathrm{Pref}(m)=
\frac{1}{NU}
\sum_{i=1}^{N}
\sum_{u=1}^{U}
\mathbbm{1}\!\left[w_{i,u}=m\right]
\times 100\%,
\]
where $N=50$ is the number of prompt-level result groups, $U=30$ is the number of participants, and $w_{i,u}$ denotes the method selected by participant $u$ for prompt $i$. Thus, each duration contains $1500$ total votes.

The exact instruction shown to participants is:

\begin{quote}
\small
You will see one text prompt and multiple anonymized videos generated from the same prompt. The model names are hidden, and the video order is randomized. Please compare all videos and select the single best video by jointly considering overall visual quality, temporal consistency, motion naturalness, and caption alignment.

For 30-second trials, please watch each video in full before making your selection. For 240-second trials, please watch at least 60 seconds in total from each video. In particular, the first 10 seconds and the final 10 seconds, i.e., 230--240 seconds, must be watched in full. For the middle part, i.e., 10--230 seconds, you may freely use the progress bar to inspect segments of interest, but please watch at least 40 additional seconds before making your selection. Please pay attention to long-range subject identity, background consistency, motion coherence, and whether the video remains aligned with the text prompt.
\end{quote}

The evaluation interface shows only the text prompt and anonymized videos in randomized order. The preference records used for analysis contain only anonymous forced-choice selections and do not include personal or sensitive information. If administrative information is required for compensation, it is handled separately from the evaluation interface and is not linked to the anonymous preference records.

\section{Limitations and Future Works.}
\label{appendix:limitations}

\ourmethod{} has several limitations.
(1) The memory queries and unified relative RoPE recipe are designed to improve quality and consistency for long/infinite video generation. Optimizing them specifically for interactive video generation remains open.
(2) Limited by the base model scale and generative capacity, \ourmethod{} may show reduced stability for hour-scale or longer videos with highly dynamic scenes.
(3) Quantitative evaluation of ultra-long videos remains an open challenge. Existing video-generation benchmarks and metrics~\citep{huang2023vbench,huang2025vbench++,zheng2025vbench2} are mainly designed for short- or minute-scale videos, and do not yet provide targeted and efficient measurements for hour-scale or day-scale temporal consistency, such as identity drift, layout drift, repetition, and prompt forgetting. Since designing such an evaluation suite is orthogonal to our memory-learning framework, we complement existing 30s/240s quantitative evaluations with extensive qualitative ultra-long results on our project page, and leave principled quantitative ultra-long evaluation to future work.
For future work, we believe:
(1) Studying the semantics of memory queries could enable controllable retrieval-augmented video generation.
(2) Distilling \ourmethod{} into a one-step model could improve real-time throughput while preserving long-term consistency.
(3) Developing efficient and targeted metrics for ultra-long video generation could better quantify hour-scale and day-scale consistency.

\section{Broader Impacts}
\label{appendix:broader_impacts}

\paragraph{Positive impacts.}
\ourmethod{} contributes two core designs to long-video generation: end-to-end learnable \memoryquery{} that filter, abstract, and compress arbitrary-length history with constant per-step cost, and a unified \relativerope{} recipe that aligns the training and inference distributions of temporal RoPE indices.
Together, they support coherent long-video generation up to hour-scale in real time, enabling applications such as long-form storytelling, immersive education, accessibility tooling, and interactive content creation.

\paragraph{Negative impacts.}
Like other capable video generation models, \ourmethod{} could be misused for disinformation, non-consensual likeness generation, or impersonation; its ability to produce coherent multi-hour videos without quality degradation may further amplify the scale of such misuse.

\paragraph{Mitigation.}
We recommend pairing any public release with provenance signals such as watermarks that interoperate with current synthetic-media detectors, and with usage policies that prohibit non-consensual personal-identity generation.
Our training relies only on publicly released prompt collections (VidProM~\citep{vidprom} and MovieGen~\citep{moviegen}); we do not collect personally identifying data, and the model does not make decisions about specific individuals or groups.
For higher-risk deployment scenarios such as interactive long video with prompt switching, we further suggest gated or staged release with explicit misuse-monitoring channels.

\newpage
\end{document}